**Predictions and Uncertainty Estimates of Reactor Pressure Vessel Steel Embrittlement Using Machine Learning**

**Authors:** Ryan Jacobs[1], Takuya Yamamoto[2], G. Robert Odette,[2] Dane Morgan[1]

[1] Department of Materials Science and Engineering, University of Wisconsin-Madison, Madison, WI, USA

[2] Mechanical Engineering Department, University of California Santa Barbara, Santa Barbara, CA, USA.

**Abstract:** An essential aspect of extending safe operation of the world's active nuclear reactors is understanding and predicting the embrittlement that occurs in the steels that make up the Reactor pressure vessel (RPV). In this work we integrate state of the art machine learning methods using ensembles of neural networks with unprecedented data collection and integration to develop a new model for RPV steel embrittlement. The new model has multiple improvements over previous machine learning and hand-tuned efforts, including greater accuracy (e.g., at high-fluence relevant for extending the life of present reactors), wider domain of applicability (e.g., including a wide-range of compositions), uncertainty quantification, and online accessibility for easy use by the community. These improvements provide a model with significant new capabilities, including the ability to easily and accurately explore compositions, flux, and fluence effects on RPV steel embrittlement for the first time. Furthermore, our detailed comparisons show our approach improves on the leading American Society for Testing and Materials (ASTM) E900-15 standard model for RPV embrittlement on every metric we assessed, demonstrating the efficacy of machine learning approaches for this type of highly demanding materials property prediction.

# 1.    Introduction:

Nuclear power is a key component of global clean energy production, producing roughly 20% of the power in the US and 25% of the power in the EU as of 2021. For perspective, as of 2019 this represents approximately half of the low-carbon electricity generation in the EU.[1,2] Given the urgent need to shift from fossil fuels to green energy sources to mitigate the numerous negative effects of global climate change,[3] the role of nuclear energy as a source of clean electricity will be even more important in the future. In the near term, nuclear energy's contribution will depend on life extension of the existing fleet of reactors.

A key component of light water nuclear reactors (LWRs) is the massive heavy section reactor pressure vessel (RPV), which houses the nuclear fuel core and pressurized reactor coolant. The RPV is fabricated from low-alloy steels. Since RPVs are individually custom-built, they contain a wide range of Cu, Ni, Mn, P, Si, C and other dissolved solutes. During reactor operation, the neutron flux leaking from the reactor core impinges the walls of the RPV, resulting in embrittlement of the steel, manifested as an upward shift in the ductile to brittle transition temperature (TTS). Embrittlement is mainly the result of radiation-enhanced formation of Cu-rich and Mn-Ni-Si-rich precipitates (CRPs and MNPs).[4–6] The precipitates impede dislocation glide, resulting in irradiation hardening manifested as an increase in the yield stress ($\Delta\sigma_y$), which in turn results in an increase of the TTS. Replacement of the RPV is not feasible. Thus, it is essential to understand, and accurately predict, how the RPV steels embrittle as a function of the neutron fluence (with typical units of neutrons $n/cm^2$), which is the corresponding flux ($n/cm^2$-s) integrated over time, up to 80 years, or more, of extended reactor life. Note, a detailed review of embrittlement mechanisms and associated predictive TTS models can be found in the review from Odette et al.,[6] hence, there is no attempt to cover the many details therein in this machine learning focused paper.

There are two main sources of embrittlement data. The first and most relevant is from plant-specific surveillance programs. A set of surveillance capsules, mounted on the inner vessel wall, contain test samples what were believed to be the most embrittlement-sensitive welds and base metals in the corresponding RPV. The capsules, which contain tensile and Charpy V-notch impact test specimens (and in some cases fracture toughness specimens), are periodically removed and tested to monitor embrittlement as a function of fluence. The fluence of these specimens closely tracks, though is modestly higher than, the exposure of the vessel itself. The corresponding tensile tests are used to track the evolution of the yield stress, ultimate tensile strength and ductility of the material. Impact tests, which measure the energy absorbed in breaking Charpy V-notch specimens as a function of temperature, are used to evaluate the TTS. The transition temperature is typically indexed at an absorbed energy of 41J.[5–7] The other source of embrittlement data is accelerated irradiations in test reactors, generally at significantly higher fluxes than in the surveillance database. Thus, the potential effect of flux must be considered in the interpretation and use of test reactor embrittlement data.

A number of physically-motivated, semiempirical models have been developed to predict TTS based on surveillance data.[6,8] The three most commonly used models are the E900 model (which is the ASTM standard),[9] the Eason-Odette-Nanstad-Yamamoto (EONY) model,[4] and the recent Odette-Wells-Almirall-Yamamoto (OWAY) model.[6] The E900 and EONY models consist of physically-motivated analytical functions with parameters that were fitted to surveillance data as new data became available and understanding of the underlying physics matured. The OWAY model uses surveillance data, or TTS predictions, at intermediate fluence combined with a set of slightly flux-adjusted test data at fluence somewhat greater than the peak at 80-year life, to interpolate the TTS in between low and high fluence.

Distinct from these physics-informed semiempirical models of RPV steel embrittlement are fully data-driven approaches which leverage machine learning (ML) techniques. The use of ML techniques in materials science has exploded in recent years.[10–14] ML models differ from non-data-driven approaches because they do not rely on fitting multiple parameters for complex, ideally physical, models to experimental data. Thus, ML can significantly reduce the time and materials domain-specific expertise needed to extract empirical information. In addition, many ML models allow for determination of uncertainty estimates on new predictions, and can provide additional insight on the effects of variable combinations not represented in current physics-based models. However, the RPV ML models generally have little or no physical constraints and can be difficult to interpret (an issue sometimes captured by referring to the models as being a "black-box"), which can lead to lack of trust in predictions. To balance the lack of physical constraints and black box nature of ML models, it is necessary for them to be rigorously assessed to verify that the model produces acceptably accurate predictions when data is available for comparison, that the predictions follow known physical trends, and that the model uncertainty estimates are trustworthy. The development and validation of accurate ML models to predict RPV embrittlement would be especially useful for informing 80-to-100-year life extension conditions, associated with low flux (e.g., $3\times10^{10}$ n/cm$^2$-s) and relatively high fluence (e.g., approximately $10^{20}$ n/cm$^2$ for 100-year life extension) conditions, which are sparsely available, potentially requiring extrapolation from accelerated high flux, relatively high fluence experiments. It is worth noting that the exact fluence incurred by a reactor at 100-year life extension will vary based on the reactor type and operational conditions and may vary in the approximate range of mid-$10^{19}$ to $10^{20}$ n/cm$^2$. We use the value of $10^{20}$ n/cm$^2$ as an approximate upper bound.

There have been a number of data-centric ML studies aimed at predicting embrittlement of irradiated steel alloys over the past 15 years. These studies include prediction of irradiated yield stress and TTS of reduced activation ferritic/martensitic (RAFM) alloys for use as materials in high dose applications like fusion plants,[15–18] as well as $\Delta\sigma_y$ and TTS prediction for RPV steel alloys.[19–28] A detailed summary of these previous studies is given in the recent review from Morgan et al.[29] Here, we summarize key findings of the three studies most relevant for the present work. First, the work of Liu et al.[24] used kernel ridge regression models to predict $\Delta\sigma_y$ for the Irradiation Variable (IVAR), Belgian Reactor 2 (BR2) and Advanced Test Reactor 1 (ATR1) data comprising 1501 data entries, which they refer to as "IVAR+" in their work, and which is included in what we refer to as the "UCSB" dataset in this work (see **Section 2** for more information on the datasets used in this work). They fit $\Delta\sigma_y$ to a 9-element feature vector consisting of atomic fractions of Cu, Ni, Mn, Si, P, and C, temperature, flux and effective fluence and performed a suite of detailed random and targeted grouped cross validation (CV) tests. From random 5-fold CV, they obtained a low root-mean-squared error (RMSE) of 14.7 MPa, and RMSE of at most 25.5 MPa from the more demanding grouped CV tests. Given that the experimental uncertainty of $\Delta\sigma_y$ is about 15 MPa, the work of Liu et al. demonstrated that kernel ridge-based ML models can accurately predict RPV embrittlement, at least for the IVAR+ data considered in their work. Second, work by Ferreño et al.[25] assessed multiple ML models for predicting TTS values in the PLOTTER-15 surveillance database, comprising 1878 data entries. They trained their models considering the weight fractions of Cu, Mn, Ni, and P, temperature, flux, fluence, and the categorical features of product form (e.g., plate, weld, forging). Ferreño et al. found that gradient boosting models resulted in the best fits, with a test data RMSE of 10.5 ˚C from a single split. This result is about 20% lower than the E900 model RMSE of 13.3 ˚C, indicating improved prediction statistics and further demonstrated the promise of ML RPV embrittlement models. Finally, the work of Mathew et al.[22] used a Bayesian neural network model to predict TTS using a combination of an initial version of the IVAR test reactor database and the US NRC embrittlement surveillance database (they refer to this database as "NUREG" in their work and which forms a portion of the PLOTTER surveillance database) used in developing the EONY model. To fit their ML model, they convert the $\Delta\sigma_Y$ IVAR values to TTS using a constant factor of 0.6 ˚C/MPa, and used an 8-element feature vector comprising weight fractions of Cu, Ni, Mn, Si and P, temperature, and flux and fluence raised to the 1/2 power. They report IVAR and NUREG average $\Delta\sigma_Y$ errors of 16.1 and 30.7 MPa, respectively, which, using the 0.6 ˚C/MPa conversion, corresponds to 9.7 and 18.4 ˚C, respectively. These errors are higher than our present ML model results on UCSB (an updated and expanded version of the IVAR data used by Mathew et al.) and PLOTTER (an updated version of the EONY database). The use of Bayesian neural networks enabled Mathew et al. to also report error bars on their predicted values, though no analysis was present to assess their accuracy.

Based on the results of Liu et al.,[24] Ferreño et al.,[25] Mathew et al.,[22] and other studies,[19–21,23,26,30] a variety of ML (e.g., neural network, gradient boosting, kernel ridge regression, depending on the study) models have been successfully fit to $\Delta\sigma_Y$ and TTS RPV embrittlement data. These fits have been conducted on various sets of RPV embrittlement data both from test reactors (IVAR, BR2, ATR1) and surveillance data (PLOTTER-15), as well as the test reactor RADAMO database[20] and some combination of test reactor and surveillance data,[22] and these models typically show very good predictive ability of TTS values on par with the ASTM E900-15 model. While previous models show very promising random CV scores, there is uncertainty regarding the reliability of model predictions under high fluence conditions in general, and under low flux, high fluence conditions representing 80-to-100-year reactor life extension as well. Ferreño et al.[25] explicitly pointed out that one shortcoming of their model was the lack of low flux, high fluence data in the PLOTTER-15 database, potentially limiting the ability of the ML model to accurately extrapolate to life extension conditions. In addition, there has been very little work devoted to quantitatively assessing the accuracy of ML model uncertainty estimates (i.e., error bars on predictions) specifically for the application of predicting RPV steel embrittlement. Having an ML model with well-validated uncertainty estimates would foster greater confidence in new predictions, such as those for low flux-high fluence conditions. We also note that models to date have fit to a range of databases, but none have been fit to all of the available databases (e.g., no single paper uses even the union of the databases that can be found in these papers), which we attempt to do here. Finally, at present there is no accessible ML model that can be easily used by the broader community. Such access is important not only to encourage rigorous testing by domain experts, but for the continual improvement of the model as new data and ML techniques become available.

In this study, we work to address all of the issues raised above. We do so by (1) fitting ML models to the largest RPV embrittlement database ever used for ML studies to date by combining the latest versions of test reactor data from IVAR, ATR1, BR2 and ATR2 (collectively referred to as the UCSB database in this work, see **Section 2**), the latest surveillance database of PLOTTER-22, as well as test reactor data from the RADAMO[20] database and PLOTTER-15 MTR databases. (2) We use an ensemble of neural networks together with the recalibration approach of Palmer et al.[31] to produce well-validated uncertainty estimates on our model predictions. (3) Finally, we make the trained model publicly available on Github and can be run on Google Colab (see **Data and Code Availability**). Overall, our model shows improved performance compared to the ASTM E900-15 model on both PLOTTER surveillance data and UCSB test reactor data. Most notably, our model shows significantly lower error for PLOTTER high fluence data, suggesting that it provides improved predictions of high fluence surveillance data relevant to life extension of the US reactor fleet.

## 2.    Data and Methods:

Here, we describe the sources of RPV embrittlement data used in this work. The first source contains hardening data comprising the Irradiation Variables (IVAR), Belgian Reactor (BR2), and Advanced Test Reactor 1 and 2 (ATR1 and ATR2) data obtained by researchers at the University of California Santa Barbara, which we collectively refer to as the "UCSB" database throughout this work.[32] A detailed description of this database was provided in the previous work of Liu et al.,[24] and here the database has been updated to include the latest ATR2 data, resulting in a total of 1556 data entries. The UCSB database includes the composition information of Cu,

Mn, Ni, P, Si, and C for each RPV steel alloy, together with its flux, fluence, product form (i.e., plate, weld, forging, or standard reference material (SRM)) and measured hardening $\Delta\sigma_y$. More details regarding the UCSB database can be found in the works of Liu et al.[24,28] **Figure 1** shows the ranges of flux and fluence included in the UCSB database, where the pink, blue, green, and red points denote data for the ATR1, ATR2, BR2 and IVAR data groups, respectively. The ATR1 data consists of very high flux and fluence ($2 \times 10^{14}$ n/cm$^2$-s and $10^{21}$ n/cm$^2$, respectively), while the ATR2 data consists of high fluence of $\approx 1.4 \times 10^{20}$ n/cm$^2$ but at lower fluxes of $\approx 3.7 \times 10^{12}$ n/cm$^2$-s. While this fluence level is consistent with fluence of 80 to 100-year life extension in some PWRs, the flux level is still about two orders of magnitude higher than a typical commercial RPV. We note that, due to their extremely high flux and fluence values relative to the rest of the database, the ATR1 points are left out of the model fit, reducing the database by only 6 data points, but resulting in improved model fits in the flux-fluence range of interest. The BR2 data have flux levels that are high, lying between the ATR1 and ATR2 data. For BR2, the fluence levels range from $10^{19}$ n/cm$^2$ to slightly above $10^{20}$ n/cm$^2$. The IVAR data are at consistently lower flux and fluence levels, with fluxes ranging from $7 \times 10^{10}$ n/cm$^2$-s to $\approx 10^{12}$ n/cm$^2$-s, while the fluence levels range from very low (below $10^{17}$ n/cm$^2$) up to $\approx 3 \times 10^{19}$ n/cm$^2$-s. A clear limitation of the test reactor data is the lack of low flux, high fluence data approaching the 40 to 100-year life extension conditions. A summary of detailed statistics of the UCSB database, split out by data group, is provided in **Table S1** of the **SI**.

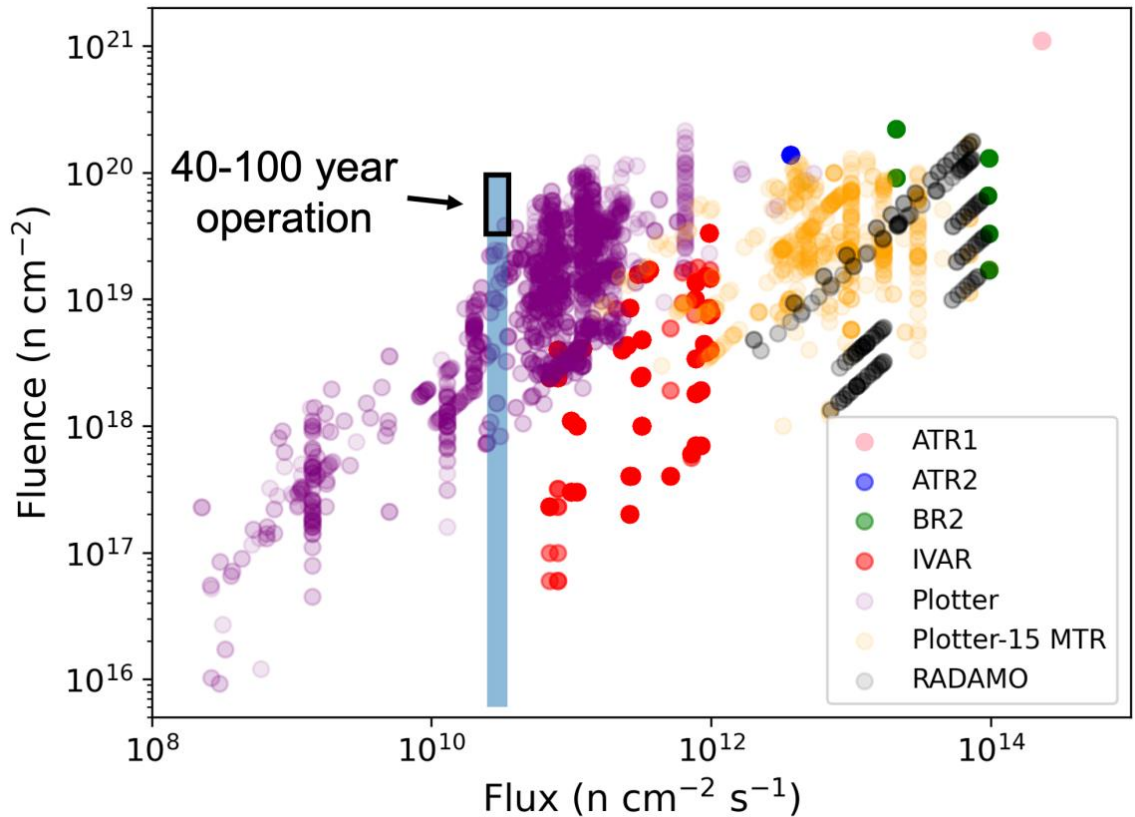

**Figure 1.** Summary of the flux and fluence values of RPV embrittlement data evaluated in this work. The colors denote different data sub-groups. The blue bar with labeled black box denotes the approximate operation conditions of 40 to 100-year life extension.

The second source of data used in this work is the PLOTTER-22 database. The PLOTTER database comes from LWR surveillance programs, collected from technical reports of individual plants from a number of countries. An earlier version, PLOTTER-15, was used to develop the E900 ASTM standard model.[9,25] The PLOTTER-22 database used in this work is an updated version of the PLOTTER-15 database, was also used in the work of Ferreño et al.,[25] and contains a total of 2048 entries (the previous PLOTTER-15 version contained 1878 entries). Similar to the UCSB dataset, the PLOTTER-22 database generally includes the composition information of Cu, Mn, Ni, P, Si, and C for each RPV steel alloy, together with its flux, fluence, product form (i.e., plate, weld, forging, or SRM) and measured embrittlement in the form of TTS values from Charpy V-notch impact tests. It is worth noting that unirradiated and irradiated yield strength information is present in the technical reports used to make the PLOTTER database, so values of $\Delta\sigma_y$ could, in principle, also be used. However, at the time of this writing, the PLOTTER-22 database contains only select entries of unirradiated yield strength from the work of Erickson and Kirk.[33] **Figure 1** shows the distribution of flux and fluence values for the PLOTTER-22 database, as purple points. The PLOTTER-22 data covers a wide range of flux and fluence values, with fluxes ranging from the very low mid-$10^8$ n/cm$^2$-s, with essentially all data below a flux of $10^{12}$ n/cm$^2$-s but with the higher fluxes overlapping the IVAR and ATR2 portions of the UCSB database, with a couple data points going up to mid-$10^{12}$ n/cm$^2$-s. The corresponding fluence levels range from roughly $10^{16}$ n/cm$^2$ up to $2\times10^{20}$ n/cm$^2$. Some data in the PLOTTER-22 database are close to the low flux, high fluence life extension conditions, with select data overlapping the approximate 40-year condition of $\approx3\times10^{10}$ n/cm$^2$-s flux and $\approx4\times10^{19}$ n/cm$^2$ fluence, the value of which can depend on the exact reactor and its operating conditions. A summary of detailed statistics of the PLOTTER-22 database is provided in **Table S1** of the **SI**.

We also include additional test reactor data from the RADAMO database (black points in **Figure 1**) and test reactor data present in the PLOTTER-15 database, which we denote as "PLOTTER-15 MTR" (orange points in **Figure 1**). Both the RADAMO and PLOTTER-15 MTR data tend to have higher fluence levels consistent with the upper end of the PLOTTER-22 and IVAR fluence levels, but again at higher flux levels more consistent with the BR2 and ATR2 test reactor data. The RADAMO hardening data takes the form of $\Delta\sigma_Y$ from tensile tests, while all but 15 of the PLOTTER-15 MTR points are TTS from Charpy impact tests.

As discussed above, the UCSB database, RADAMO database, and a small portion of the PLOTTER-15 MTR database contains hardening in the form of $\Delta\sigma_Y$ values for each RPV data entry, while the PLOTTER-22 and nearly all of the PLOTTER-15 MTR databases contain TTS values. Therefore, in order to combine these databases for the purposes of constructing ML models of TTS, we need to convert the $\Delta\sigma_Y$ values into TTS values. We use TTS values in this work because TTS is used in the ASTM E900-15 and EONY models, and is the basis for embrittlement regulations. The relationship between $\Delta\sigma_Y$ and TTS is dependent on alloy composition as well as flux and fluence. Detailed models to predict the conversion factor for a given alloy and irradiation condition are presently being developed.[34] The previous work of Mathew et al.[22] made the assumption that there is a linear relationship between $\Delta\sigma_Y$ and TTS as: TTS = $cc\times\Delta\sigma_Y$, where $\Delta\sigma_Y$ has units of MPa, and TTS has units of ˚C, and $cc$ is the conversion coefficient. Mathew et al. used a *cc* value of 0.6 ˚C/MPa.[22] While they did not cite a reference for this value, our independent analysis of the literature, described below, yielded a value of *cc* = 0.61 ˚C/MPa. In this work, data of instances where both $\Delta\sigma_y$ and TTS were obtained for the same material at a given irradiation condition were collected from the literature. A total of 311 data points were collected from the works of Odette et al.,[35] Lee et al.,[36] and Nanstad et al.[37] The WebPlotDigitizer tool was used to extract the data available only in graphical form. Of the 311 total collected data points, 8 were flagged and removed as outliers based on their ratio of TTS/$\Delta\sigma_y$ either being negative or greater than 10 (suggesting totally unphysical behavior), leaving 303 points. These points, which are plotted along with a quadratic fit, are shown in **Figure 2**. We used this fitted quadratic form to convert all of the $\Delta\sigma_y$ values in the various databases to TTS values.

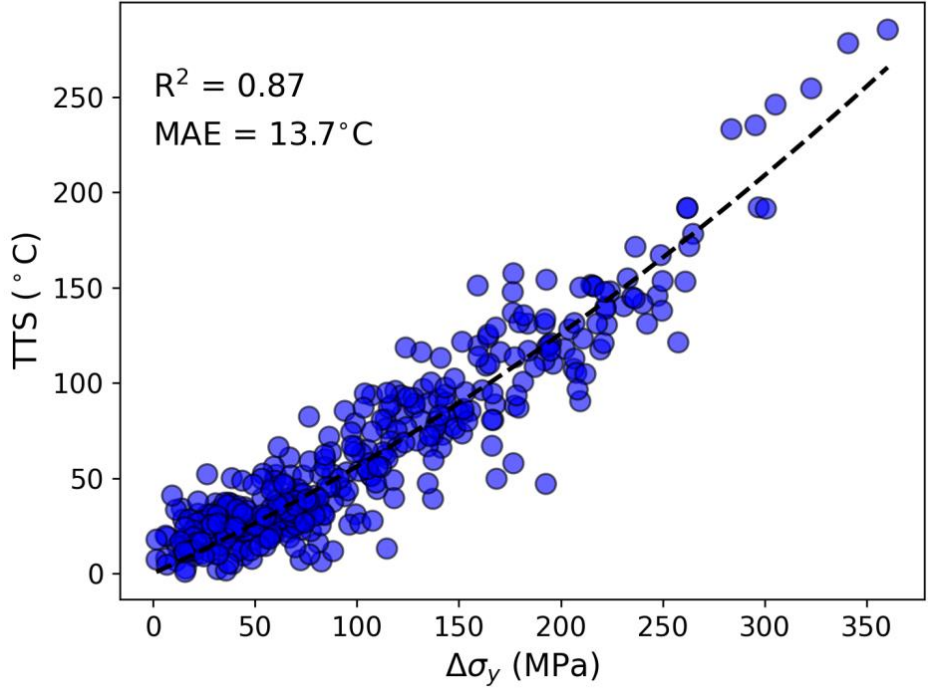

**Figure 2.** Experimental data of transition temperature shift (TTS) vs. $\Delta\sigma_y$ for 303 measurements. The best-fit quadratic equation is TTS = $0.00067\Delta\sigma_y^2 + 0.49\Delta\sigma_y$.

All ML model fits and assessments in this work were performed using the MAterials Simulation Toolkit for Machine Learning (MAST-ML).[38] Briefly, MAST-ML simplifies and accelerates the use of machine learning models in materials science, enabling users to quickly perform assessments of multiple models and cross validation tests, with codified analysis data and figure output. MAST-ML leverages widely used ML libraries like scikit-learn[39] and Keras[40] (Tensorflow[41] backend). To evaluate our ML models, we perform three types of tests. The first test is referred to as a "full fit", which fits to all of the data. A full fit is useful for deployment of a final model, however, due to the potential for overfitting, full fit errors may not represent the typical model error on new test data. The second test is random cross validation, where random subsets of the data (here, 20%) are used as test data while the remaining 80% is used as training. This test provides a reliable assessment of model performance on test data that is similar to (i.e., within the domain) of the training data. The third test is leave out group cross validation, where the test data is an entire distinct group of data (e.g., PLOTTER-22 surveillance data), and the training data is everything else (e.g., UCSB, RADAMO, PLOTTER-15 MTR). This test is more demanding than random CV and is meant to mimic how the model may perform when given test data that is significantly different (i.e., partially or totally outside the domain) from the training data. MAST-ML also includes a robust uncertainty quantification framework with error bar recalibration using the methods of Palmer et al.[31] In this work, we focus on using an ensemble of fully-connected neural networks as our model. We construct an ensemble of 10 fully connected neural networks using the *EnsembleModel* class in MAST-ML, which builds upon scikit-learn's *BaggingRegressor* class. Initial tests of 1, 2, 5, 10, and 15 networks in the ensemble saw saturation of performance by 10 networks. For all fits, the features used are Cu, Ni, Mn, P, Si and C content (in weight percent), temperature, and the base-10 logarithm of flux and fluence, for a total of 9 features. For PLOTTER-22 entries that contain blank entries for Si or C data, values of 0 wt% were used. In **Figure S1** of the **SI**, we plot the true (i.e., measured) TTS values as a function of each feature.

Each network in the ensemble was fit using all 9 features in the dataset, and the different networks are fit on bootstrapped samples of the training data. Approximately 25 different model architectures were manually constructed and tested to find a model which resulted in low 5-fold CV RMSE while not being too computationally expensive. This search of model architecture is by no means exhaustive, and while a more ideal model architecture may exist, the present model was found to give accurate full fit and 5-fold CV predictions, have well-calibrated uncertainty estimates, and demonstrate reasonable prediction trends with known physics. The neural network used in the present work was built using Keras, run in the MAST-ML package, and consists of a sequential (fully-connected) neural network with 5 layers. The first layer has an input dimension of 9 and is a dense layer with 1024 nodes. The second layer is a dropout layer with dropout rate equal to 0.3. The third layer is another 1024-node dense layer. The fourth layer is another dropout layer with dropout rate of 0.3. The fifth layer is the single-node prediction layer. The rectified linear activation function was used throughout, and the model was optimized using the Adam optimizer with mean squared error as the metric.

Similar to work from Liu et al.,[24] we added artificial data points to our fits that, for each alloy, correspond to a TTS = 0 ˚C value for fluence of $10^{15}$ and $10^{16}$ n/cm$^2$ at fluxes of $10^8$, $10^9$, $10^{10}$, $10^{11}$ and $10^{12}$ n/cm$^2$-s, resulting in a total of 980 anchor points for the UCSB database (98 alloys x 2 fluences x 5 fluxes) 6240 anchor points for the PLOTTER database (624 alloys x 2 fluences x 5 fluxes), 100 anchor points for the RADAMO database (10 alloys x 2 fluences x 5 fluxes), and 2040 anchor points for the PLOTTER-15 MTR database (204 alloys x 2 fluences x 5 fluxes). These anchor values were used as training data for every data split but were not used as test data and thus were not used in computing any model performance metrics. The purpose of including these data is to provide the model with additional, physically-motivated anchor points which better enable the model to predict zero embrittlement at low fluence for a wide range of flux values.

# 3.    Results and Discussion:

## 3.1    Machine learning fits on individual databases

In this section, we first evaluate the performance of our ML model for predicting TTS for the UCSB and PLOTTER databases individually. We focus on these two databases for several reasons. First, UCSB and PLOTTER represent large test reactor and surveillance databases, respectively, allowing comparison across these two different classes of data. Second, within our sets of test reactor data, UCSB was based on controlled single and combined variable experiments resulting in a consistent, high-resolution map of embrittlement dependencies, covering a very wide range of alloy and irradiation conditions. This consistency minimizes contributions from non-physical factors in assessing the model. Finally, the ML predictions on these two data sets can be readily compared to previous research. Then, in **Section 3.2**, we assess performance of our ML model on the combined database. Here, we evaluate the performance of our ML model based on full fits to all of the data, and random 5-fold CV. **Figure 3** contains parity plots assessing these fits, and **Table 1** contains a summary of the per-group RMSE values for all fits. **Figure 3A** and **Figure 3B** show the full fit results on the UCSB and PLOTTER databases, respectively. Our model shows very low full fit RMSE values of 5.2 ˚C and 8.0 ˚C for UCSB and PLOTTER, respectively. The RMSE of 5.2 ˚C on UCSB data is on par or slightly better than the value obtained by Liu et al. (note they fit to $\Delta\sigma_y$ in their study).[24] In addition, the RMSE of 8.0 ˚C on the PLOTTER data is within the experimental uncertainty of the TTS values, which is approximately 10 ˚C, and was quoted to be 7-8.5 ˚C by Ferreño et al.[9,25] **Figure 3C** and **Figure 3D** show the 5-fold CV results on the UCSB and PLOTTER databases, respectively. Our model shows broadly good predictive ability, with average RMSE values of 9.2 +/- 1.4 ˚C and 11.9 +/- 0.8 ˚C across all folds on UCSB and PLOTTER, respectively (note that the +/- values are standard deviation in the RMSE based on data from across 25 random folds). Our average CV RMSE on PLOTTER data of 11.9 ˚C is slightly higher than the RMSE of ≈10.5 ˚C on a single test split obtained by Ferreño et al., who used a tuned gradient boosting model. We note here that 10.5 ˚C is within our cross-validation sampling error bar, indicating some of our data splits obtain this level of performance.[25] While the exact data, feature set, and model type differ between this work and Ferreño et al., both of these RMSE values on PLOTTER data are better than the ASTM E900-15 model, which obtains a full fit RMSE of 13.4 ˚C (see **Section 3.4**).

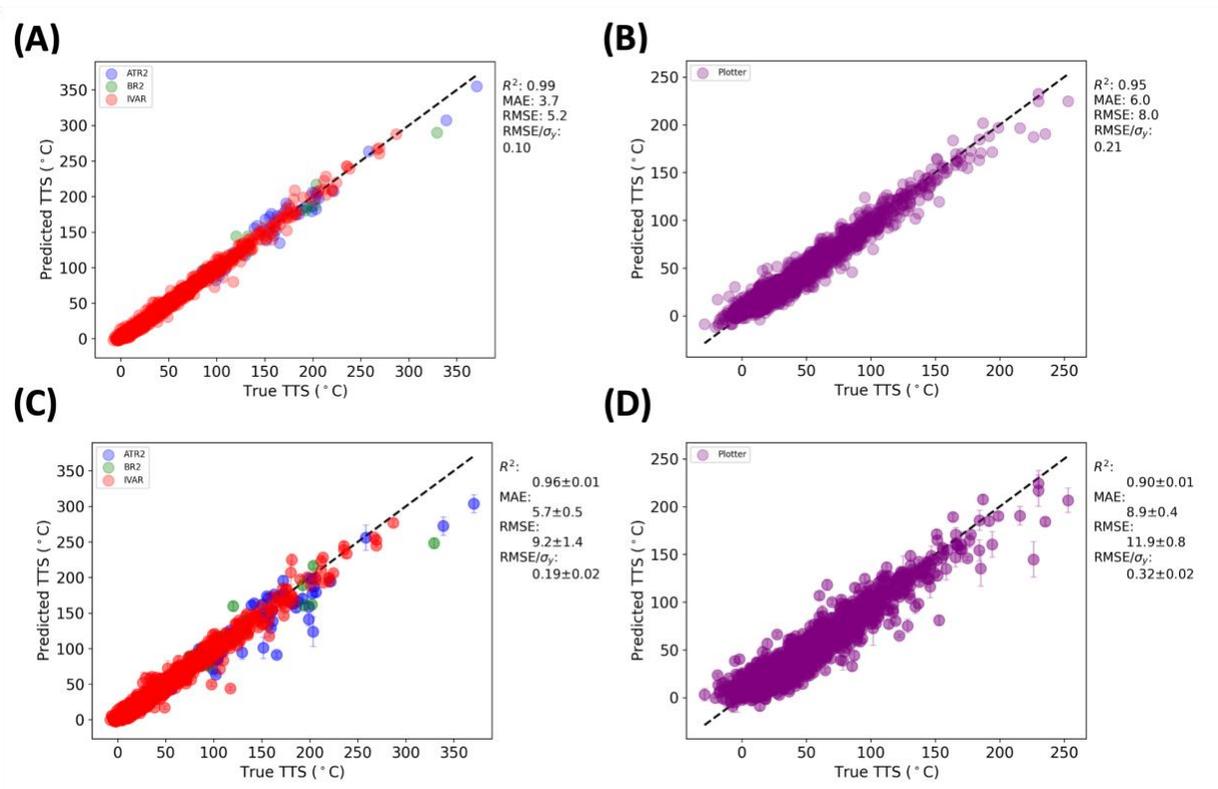

**Figure 3.** Parity plots of ML model predictions for full fits to (A) UCSB and (B) PLOTTER databases in isolation. Panels (C) and (D) correspond to ML model predictions for 5-fold cross validation for UCSB and PLOTTER, respectively. For the 5-fold CV cases, the metric values are averages +/- standard deviation over 25 splits.

## 3.2    Machine learning fits on combined databases

Here, we assess the quality of our ML model when the UCSB and PLOTTER databases are joined together into a single database, with the RADAMO and PLOTTER-15 MTR data included as well (see **Section 2** for details on how the databases were joined). **Figure 4** contains parity plots assessing our ML model on a full fit to the entire database (**Figure 4A**), random 5-fold CV (**Figure 4B**), and the more demanding leave out group CV (**Figure 4C**). **Table 1** contains a summary of the per-group RMSE values for all fits. For the full fit test, the RMSE of 8.4 ˚C is slightly higher than the full fit RMSE values of the individual datasets from **Figure 3**, which is reasonable considering the higher error on the PLOTTER-15 MTR subset and slightly higher BR2 errors in the combined database fit. Similarly, the 5-fold CV RMSE of 12.2 +/- 0.7 ˚C as shown in **Figure 4B** is within the error bars of the individual PLOTTER database 5-fold CV RMSE value and slightly higher than the individual UCSB database 5-fold CV RMSE value. The leave out group CV test in **Figure 4C** shows a significantly higher RMSE of 29.3 °C (note this is averaged over each left-out group). Interestingly, the PLOTTER data is still predicted quite well for this test, with an RMSE of 18.2°C, an impressive result considering no PLOTTER data was used in training for that data split. In this case, we speculate the robustness of our model predictions on PLOTTER data is due to the inclusion of the IVAR data, which covers similar, though smaller, flux and fluence ranges compared to PLOTTER. We discuss the merits of our ML model on high fluence data in **Section 3.3**. In **Figure S2** of the **SI**, we plot the TTS residuals as a function of each input feature. Generally, we find that the residuals are highly symmetric about zero, and there is no appreciable trend in the residuals with any feature. This is an important aspect of the ML model predictions, as any trend of the residuals with, for example, fluence, would constitute undesirable behavior of the model.

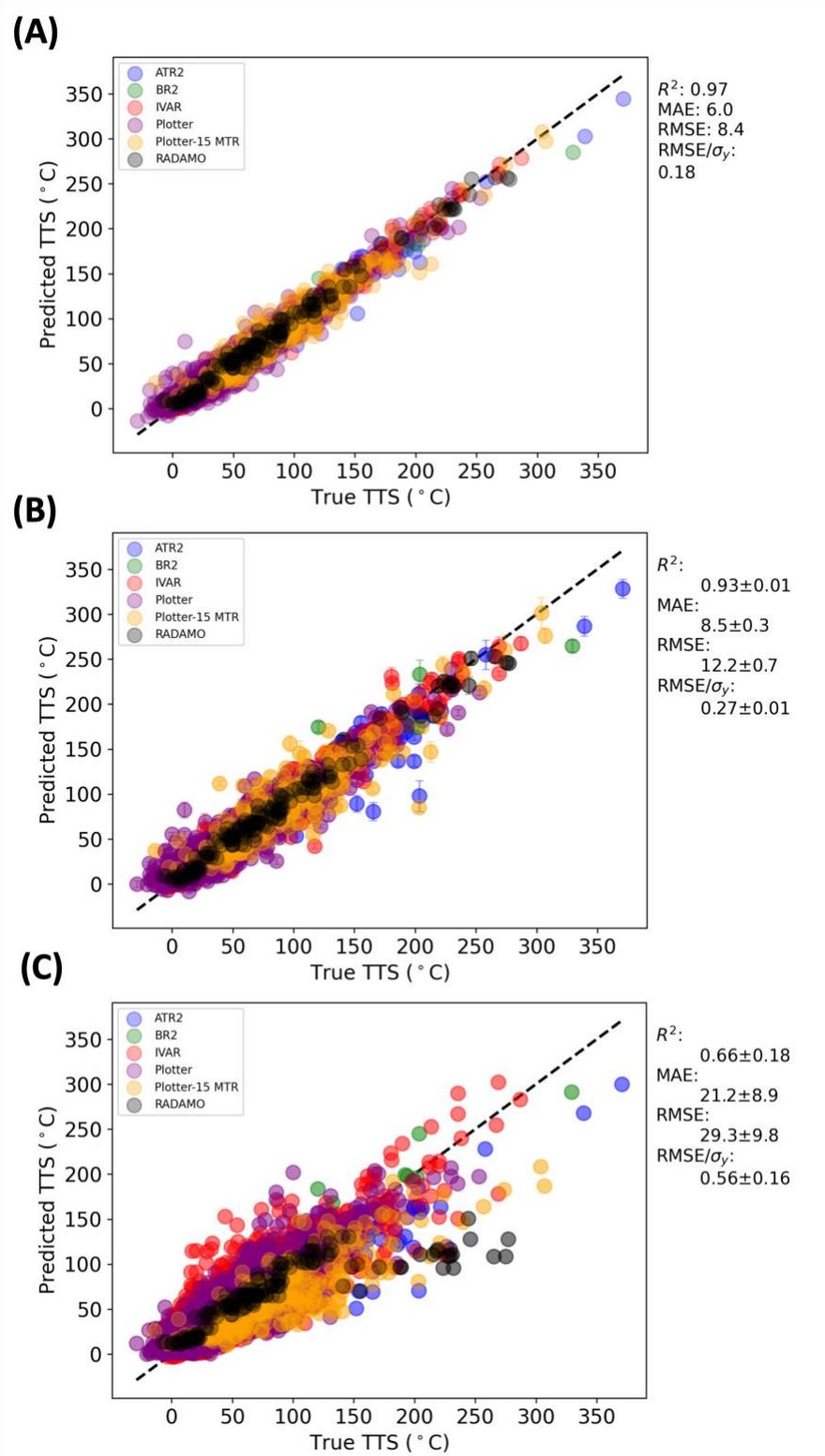

**Figure 4.** Parity plots of ML model predictions on the combined UCSB+PLOTTER database. (A) Full fit to all of the data. (B) 5-fold cross validation results. (C) Leave-out-group cross validation. The groups of ATR2, BR2, IVAR, PLOTTER, PLOTTER-15 MTR and RADAMO are shown as blue, green, red, purple, orange, and black points, respectively.

**Table 1.** Summary of ML model RMSE values, broken out by data group, comparing performance of full fit and 5-fold CV tests when the models were trained on the UCSB and PLOTTER databases in isolation vs. when the combined database was used.

| Data group | Full fit RMSE (individual dataset) | Full fit RMSE (combined dataset) | 5-fold RMSE (individual dataset) | 5-fold RMSE (combined dataset) | Leave out group RMSE (combined dataset) |
|---|---|---|---|---|---|
| Plotter | 8.0 | 8.4 | 12.0 | 11.6 | 18.2 |
| IVAR | 4.5 | 5.2 | 7.1 | 8.0 | 22.0 |
| BR2 | 11.5 | 11.7 | 21.0 | 21.9 | 22.5 |
| ATR2 | 12.4 | 12.9 | 30.1 | 32.3 | 47.6 |
| RADAMO | n/a | 6.5 | n/a | 8.4 | 32.7 |
| Plotter-15 MTR | n/a | 13.0 | n/a | 18.9 | 32.5 |

## 3.3    Assessments of model predictions for high fluence data

Developing ML models to accurately predict high fluence, extended-life embrittlement is a primary objective of this study. In this section, we discuss the ability of our ML model to predict TTS of high fluence embrittlement. Here, we again evaluate our ML model for full and random 5-fold CV fits. We then assess the role of low fluence data on the high fluence predictions. This involves comparing fits to only the high fluence data subset to fits for all fluences. We define "high fluence" as data at, or above, a fluence of $6 \times 10^{19}$ n/cm$^2$; all other data is categorized as "low fluence". **Figure 5** shows parity plot assessments of our ML model fit and tested on high fluence data only (**Figure 5A** and **Figure 5B**), where **Figure 5A** is a full fit to all high fluence data and **Figure 5B** is 5-fold CV on high fluence data. As expected, the full fit RMSE on high fluence data of 15.5 ˚C is lower than the corresponding 5-fold CV RMSE of 25.3 +/- 2.7 ˚C. **Figure 5C** and **Figure 5D** compares these fits to those using all of the data, and the corresponding high fluence

RMSE values are summarized in **Table 2**. The use of low fluence data in fitting resulted in a reduction in the high fluence RMSE, with a full-fit (5-fold CV) RMSE reduction of 3.4 (3.8) ˚C, respectively, which is about a 22 % (15 %) decrease.

The most important comparison is not the RMSE on *all* high fluence data, but rather the RMSE on the high fluence subset of PLOTTER surveillance data most representative of RPV service conditions. These RMSE values are summarized in **Table 2**. Here, it is evident that inclusion of low fluence data significantly reduces the error on high fluence PLOTTER data, for both the full fit and 5-fold CV cases, with an RMSE reduction from 13.6 ˚C to 10.6 ˚C for the full fit case (a 22.5% reduction) and RMSE reduction from 20.7 ˚C to 15.2 ˚C for 5-fold CV (a nearly 27% reduction). While the RMSE of the ASTM E900-15 model is 13.4 ˚C on all of the PLOTTER data, that error increases to 17.8 ˚C when considering only the high fluence subset of the PLOTTER data. Thus, both the full fit and 5-fold CV ML model RMSE values are lower than the E900 RMSE on high fluence PLOTTER data, demonstrating that our ML model offers at least comparable, if not improved, predictive ability compared to the ASTM E900 model.

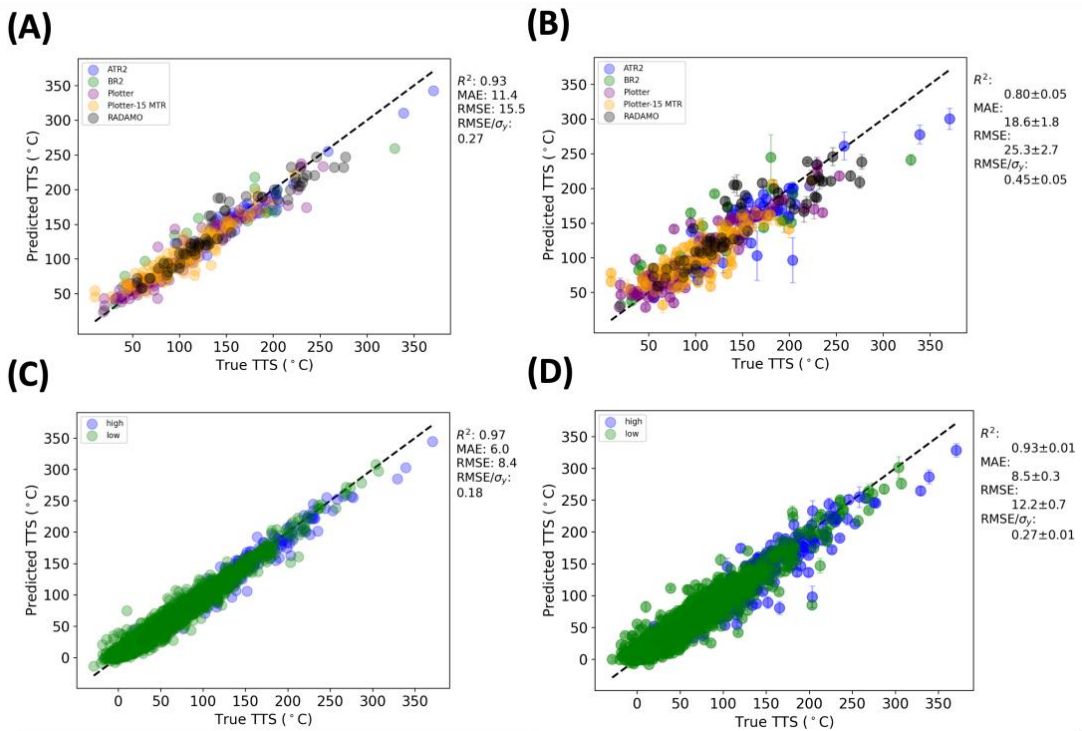

**Figure 5.** Parity plots of ML model predictions on high fluence data, where high fluence is designated as all data points with fluence values of $6 \times 10^{19}$ n/cm$^2$ and higher. Panels (A) and (B) correspond to full fit and 5-fold CV results where the data trained and tested on the subset of high fluence data only (388 data points). Panels (C) and (D) correspond to full fit and 5-fold CV results where the full database is used. In panels (A) and (B), the RMSE values on high fluence data only are 15.5 and 25.3 ˚C, respectively, and in panels (C) and (D), the RMSE values on high fluence data only are 12.9 and 20.5 ˚C, respectively (see **Table 2**).

**Table 2.** Summary of ML model RMSE values on high fluence data. All RMSE values have units of ˚C. As a reference, the ASTM E900 model has an RMSE of 13.4 ˚C on all PLOTTER data, and an RMSE of 17.8 ˚C on high fluence PLOTTER data (see **Section 3.4**).

| Data used in training | Full fit | 5-fold CV |
| --- | --- | --- |
| High fluence only | 15.5 (PLOTTER = 13.6) | 24.5 (PLOTTER = 20.7) |
| All data | 12.9 (PLOTTER = 10.6) | 20.5 (PLOTTER = 15.2) |

## 3.4    Comparison of ML model performance with E900 and EONY models

Here, we first briefly discuss the performance of the ASTM E900-15 model compared to the EONY model as these models are two of the most popular analytical models for predicting RPV embrittlement. Then, we present a detailed discussion of the performance of our ML model compared to the E900-15 model and the advantages and shortcomings of each model. We focus our detailed discussion of our ML model vs. the E900 model as E900 is a leading analytical model used for predicting RPV material embrittlement and informing life extension. **Figure 6A** and **Figure 6B** contain parity plots assessing the performance of the E900 and EONY models, respectively. A summary of the per-group RMSE values for these models, with a comparison to the respective RMSE values from the ML model, is provided in **Table 3**. Broadly, the E900 and EONY models show comparable performance, with overall RMSE values of 23.3 and 23.0 ˚C on the full database, respectively. E900 tends to show improved performance on the PLOTTER data compared to EONY (13.4 ˚C vs. 17.3 ˚C, which increases to 17.8 ˚C vs. 40.5 ˚C on the subset of high fluence PLOTTER data), while EONY tends to show improved performance on the UCSB data, particularly the ATR2 (55.5 °C vs. 81.1 °C) and BR2 data (44.5 °C vs. 87.0 °C). Overall, the EONY model is an older model than E900 and was fit to a smaller, earlier surveillance database and is thus expected to be less accurate than E900 for predicting TTS of surveillance data at life extension conditions.

Next, we compare the performance of the E900 and our ML model on the PLOTTER database, as this is the domain over which E900 was developed to have strong performance. E900 has an RMSE on just the PLOTTER data of 13.4 °C, which is a lower value than on the other data groups. We do not have any CV scores for E900 to compare to the CV values for our ML model in this work. However, a recent study by Ferreño et al.[42] found that the E900 model is not significantly overfit, and their cross-validation approach resulted in a leave out 25% test set RMSE about 2.3% higher than the full fit value (in their study, 13.5 °C vs. 13.2 °C). If we assume that a comparable percentage error would occur for all E900-related cross validation RMSE values, then we can estimate the 5-fold RMSE value for E900 on PLOTTER to be 13.7 °C. The full fit and 5-fold RMSE values of the ML model developed here on PLOTTER are 8.4 °C and 11.6 °C, somewhat lower than E900. Thus, when comparing average statistics on the total or PLOTTER database, our ML model does somewhat better than E900. As mentioned above, an important goal for TTS models is the ability to predict the TTS of high fluence data in the PLOTTER database. The E900 model has a full fit RMSE of 17.8 °C on the high fluence data (see **Table 3**), and an estimated 5-fold CV RMSE of 18.2 °C. These values are somewhat higher than the full fit and 5-fold ML model RMSE values on PLOTTER high fluence data of 10.6 °C and 15.2 °C, respectively (see **Table 3** and **Figure 6C**). Overall, the above results show that our ML model generally performs better than the E900 model on the target RMSE metrics for PLOTTER. We note here that while we focus on reporting and comparing RMSE values in our analysis, the mean absolute error (MAE) is also important, and we provide a table of MAE values for our model and others in the literature in **Table S3** in the **SI**.

It is also of interest to consider the model errors on a broader domain of data. First, the E900 model shows an overall RMSE on the total database of 23.3 °C, significantly higher than the full fit and 5-fold CV ML RMSE values of 8.4 °C and 12.2 °C, respectively. This result makes sense, considering the PLOTTER database was used to fit the E900 model and the full database was used to fit the ML model. A more targeted test of the domain of the E900 model is its ability to predict IVAR data, as this has significant overlap in composition and irradiation conditions with the PLOTTER data (although they certainly differ in many places as well). E900 has an RMSE of 22.6 ˚C on IVAR data, a significantly higher error than on the PLOTTER data, but still reasonably good considering no IVAR data was used to fit E900. The full and 5-fold CV ML model RMSE on IVAR are 5.3 ˚C and 8.0 ˚C, respectively, much lower than the E900 model, but this is to be expected as our ML model is fit with IVAR data. However, the E900 RMSE of 22.6 ˚C on IVAR data is comparable to how our ML model performed in the leave out group test shown in **Figure 4**, where an RMSE of 22.0 ˚C was obtained on IVAR data when the PLOTTER data was used as training data. The E900 model shows significantly higher errors on the BR2 and ATR2 data groups, suggesting that E900 may not accurately predict TTS for combined high flux and high fluence conditions. Again, this is not unexpected given the flux and fluence conditions used in the E900 training data. These results suggest that the ML model has a much wider domain of applicability than E900. Further, by re-fitting the 26 parameters of the E900 model to our full database, the RMSE is lowered from 23.3 ˚C to 17.4 ˚C. In doing this, the full database RMSE is reduced, however, the RMSE on the PLOTTER subset of data increases from 13.4 ˚C to 14.7 ˚C. This finding suggests that, in addition to the benefit of increased data, the flexibility of our ML model enables a broader domain of applicability compared to the E900 model, at least in its present form. Further modifications to the E900 model to add additional terms may further broaden its domain, but doing so requires significant human effort and expertise.

We have also plotted the full residual histograms of our ML model and the E900 and EONY models in **Figure 6D**. Note that the data shown in **Figure 6D** are for PLOTTER data only. Another histogram showing the distribution of residuals of the high fluence PLOTTER data only is shown in **Figure S3** of the **SI**. Regardless if we examine the residual distributions on all PLOTTER data or just the high fluence portion of PLOTTER data, our ML model has both the smallest mean error (i.e., smallest bias) and the smallest standard deviation of residuals. In addition, our ML model is fit to far more data, which, as discussed above, gives it a larger domain of applicability than E900.

Finally, we discuss the advantages and shortcomings of our ML model vs. the E900 model on the merits of (i) model errors, applicability domain and uncertainty estimates, (ii) prediction of physical trends, (iii) model interpretability, (iv) model evolution and re-use. Regarding (i) (model errors), as discussed above, compared to E900, our ML model generally has lower errors across all databases, and, importantly, the high fluence portion of the PLOTTER database. The level of improvement of our ML model over E900 on the total and high fluence PLOTTER database is notable, with a maximum of 1.8 ˚C and 2.6 ˚C (13.2 % and 14.5 % reductions), respectively. In addition, we found modest performance improvement for our ML model vs. E900 for data outside of PLOTTER when our ML model did not have access to that data. Therefore, we have some evidence that the improved performance of our ML model on a wider domain is primarily due to its having access to more data for fitting, although it is possible that the more flexible ML models do offer advantages over E900 in fitting a larger database. An important limitation of the E900 model is that it does not provide an estimate of uncertainty (i.e., an error bar) with a given prediction. As described in **Section 3.5**, a key advantage of using an ensemble of neural networks is it provides well-calibrated error bars on our predictions. Regarding (ii) (prediction of physical trends), as described in **Section 3.6**, our ML model shows no signs of unphysical behavior or problematic overfitting. This behavior is both a strength and weakness of our ML model vs. E900. It is a strength because our ML model appears able to capture the detailed known physical trends without assuming an explicit functional form, thus capturing correct embrittlement physics without needing deep domain expertise. However, since no physics-based functional form is used for our ML model, detailed testing of physical trend assessment is required to foster confidence in the model. Regarding (iii) (model interpretability), in favor of E900, it should be noted that it has far fewer parameters (26 + explicit functional form vs. ~1M for a single neural network in the ensemble). While there is no evidence that the extra parameters in the ML are introducing problematic overfitting, having fewer parameters is generally desirable. Furthermore, E900 has a simple explicit analytical form, while the functional behavior of the neural networks used in the ML model cannot be understood intuitively and only probed indirectly through methods such as cross plots. Regarding (iv) (model evolution and re-use), our ML model takes advantage of modern ML tools and is therefore easy to fit and update (i.e., re-fit) as new data becomes available. This stands in contrast with the E900 model, which requires significant manual re-tuning of the functional form and its underlying parameters and significant domain expertise to do so. Putting all these aspects together suggests the following. First, our and similar ML models are likely to provide notable improvements in performance vs. E900 within the domain of PLOTTER and likely even life-extension conditions on existing plants, but at the cost of increased complexity and decreased transparency. The greatest advantages of using ML models may come from the ability to quickly fit and improve them, particularly when integrating large amounts of data, and to take advantage of ML-associated technologies like the error estimation methods we have employed and discuss in **Section 3.5**.

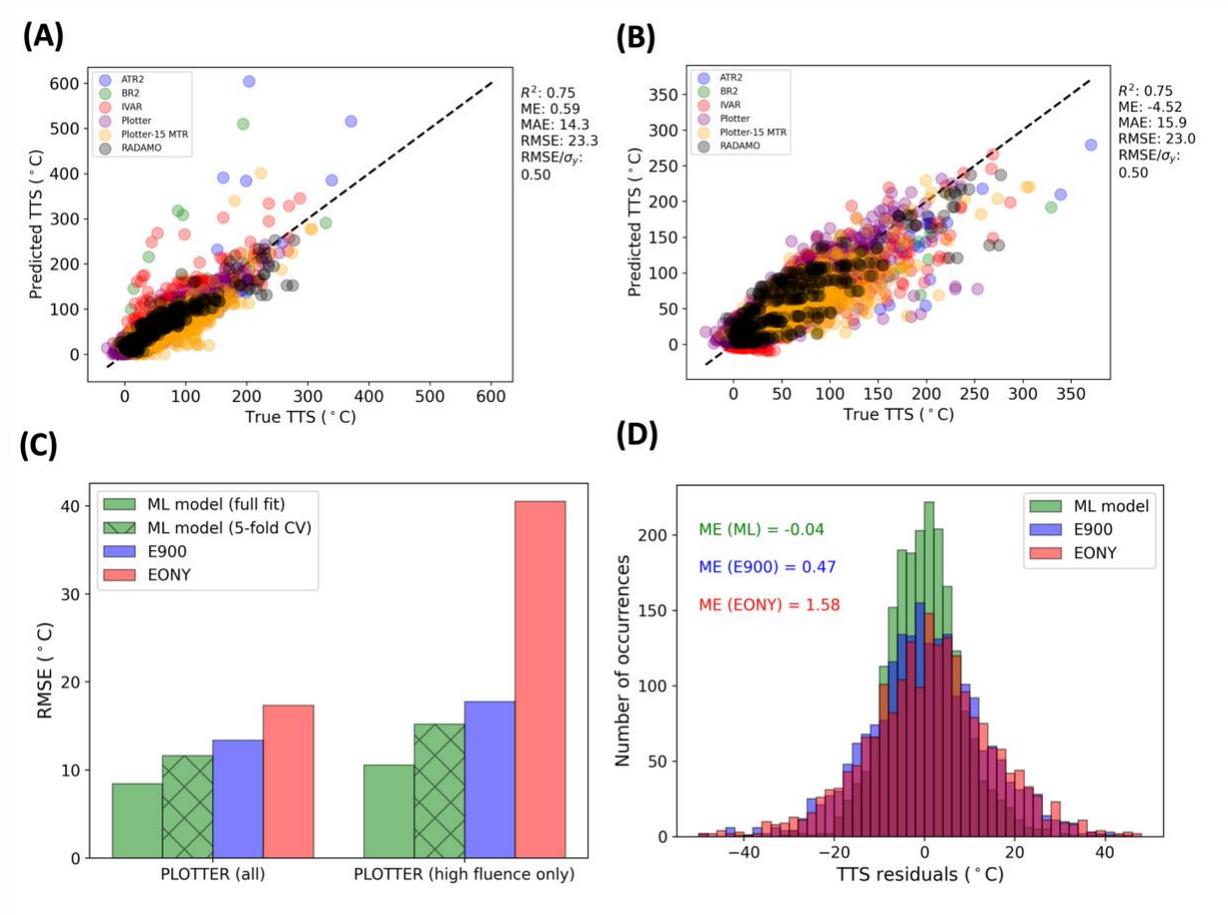

**Figure 6.** Parity plots showing predictions of the (A) E900 and (B) EONY models on all data groups of the combined database. Panel (C) is a bar chart comparing the RMSE values between our ML model, E900 and EONY on just the PLOTTER data. Panel (D) are histograms of residuals for our ML full fit model (green), E900 (blue) and EONY (red) on PLOTTER data only. The quoted error values in (D) are mean errors (ME) in ˚C. In (D), the standard deviations of the residuals are 8.4, 13.4 and 17.3 ˚C for our ML model, E900 and EONY models, respectively.

**Table 3.** Summary of RMSE values comparing E900, EONY, and our ML model for different data groups. All values are given in units of ˚C.

| Data group | ML model RMSE (full fit) | ML model RMSE (5-fold CV) | E900 RMSE | EONY RMSE |
|---|---|---|---|---|
| ATR2 | 12.9 | 32.3 | 81.1 | 55.5 |
| BR2 | 11.7 | 21.9 | 87.0 | 44.5 |
| IVAR | 5.2 | 8.0 | 22.6 | 21.5 |
| PLOTTER | 8.4 (10.6 on high fluence) | 11.6 (15.2 on high fluence) | 13.4 (17.8 on high fluence) | 17.3 (40.5 on high fluence) |
| PLOTTER-15 MTR | 13.0 | 18.9 | 30.5 | 32.2 |
| RADAMO | 6.5 | 8.4 | 23.3 | 28.2 |
| Overall | 8.4 | 12.2 | 23.3 (17.4 after re-fit) | 23.0 |

### 3.5 Assessment of ML model uncertainty quantification and well-calibrated error bars

In this section, we discuss the ability of our ML model to provide accurate uncertainty estimates (i.e., error bars) on predictions of TTS. In general, knowing that an uncertainty estimate on an ML model prediction is accurate provides confidence in the ML model prediction and better informs use of the model. Because our ML model is an ensemble of neural networks, we can obtain an uncertainty on each prediction by calculating the standard deviation of the predictions of the 10 individual neural networks comprising the ensemble. While this approach provides a simple ensemble estimate of the prediction uncertainty, one cannot tell *a priori* if this uncertainty estimate itself is accurate. We follow the approach of Palmer et al.[31] to develop calibrated uncertainty estimates and demonstrate that these uncertainty estimates are quite accurate. The work of Palmer et al. used linear rescaling to calibrate error bars. Here, we use constant, linear, and quadratic rescaling, and found that quadratic rescaling performed best. The present approach quadratically transforms the original ensemble estimate of the uncertainties so as to match the true uncertainties of test data sets. Here, we show an assessment of the raw model uncertainty estimates and the more accurate calibrated uncertainty estimates.

**Figure 7** details the assessment of ML model uncertainties both in their uncalibrated (grey data) and calibrated (blue) states. The uncalibrated values are the uncertainties directly obtained from the standard deviation of the predictions of the individual neural networks comprising our ensemble model, and are included here to show the impact of our calibration procedure and the improvement of the uncertainties after calibration is performed. In **Figure 7A**, we plot the distribution of normalized residuals, defined as the residual value divided by the corresponding model uncertainty estimate (sometimes referred to as an "r-statistic"[31,43] or a "Z-score"[44]). A well-calibrated model will have a normalized residual distribution with mean of zero and standard deviation of one, i.e., the normalized residuals will follow a unit-normal distribution. From **Figure 7A**, we can see the uncalibrated values have a mean near zero but slightly negative, and a large standard deviation of about 3.2. In contrast, after calibration, the calibrated values have mean closer to zero and standard deviation very close to 1, indicating the recalibration was successful when examining the distribution of normalized residuals. **Figure 7B** contains the same data as **Figure 7A** but plotted in the form of predicted vs. standard (true) cumulative distribution function (CDF). This plot is essentially a quantile-quantile (Q-Q) distribution plot but uses CDF instead of plotting the actual quantile values. In **Figure 7B**, the dashed line indicates a perfect unit-normal Gaussian distribution. From inspecting **Figure 7B**, it is immediately evident the calibrated distribution is much closer to the ideal distribution line than the uncalibrated distribution. The higher quality of the calibrated uncertainties is also quantitatively reflected in the calculated area values, where the uncalibrated (calibrated) data has an area of 0.114 (0.024), where an area of 0 is perfect agreement with the unit-normal distribution. Note for this plot, the area is calculated as the integral of the absolute value of the predicted CDF values and the y=x 45-degree line, which here denotes the standard normal CDF.

While the analysis shown in **Figure 7A** and **Figure 7B** demonstrate the ML model uncertainties are well-calibrated in the sense that they produce the expected distribution of values, it does not provide an indication of the direct trend between actual model errors (residuals) and their corresponding predicted errors (uncertainty estimates). To better quantify the relationship between true and predicted errors, in **Figure 7C** we make a residual vs. error (RvE) plot with quadratic rescaling of model uncertainties. RvE plots with constant, linear, and quadratic rescaling with unequal and equal sample binning are shown in **Figure S4** in the **SI** for comparison. In addition, we have plotted the distribution of normalized residuals and RvE plots with quadratic recalibration for data separated into low fluence and high fluence groups in **Figure S5** of the **SI**. This method of representing true vs. predicted model errors was introduced by Morgan and Jacobs[10] and refined by Palmer et al.[31] Briefly, residual values are binned based on the range of their uncertainty estimates. For all residual points in a given bin, the RMS of the residuals is calculated, forming one data point in the plot of **Figure 7C**. The number of residual points in each bin is reflected by the histogram overlaid above the RvE plot. In **Figure 7C**, the red dashed line is the y=x line, denoting that the predicted errors perfectly coincide with the actual model errors. Thus, a perfectly calibrated model should have an intercept of 0 and slope of 1 on the RvE plot. We can see from **Figure 7C** that the uncalibrated values have a slope of 1.25 and intercept of 0.12, indicating that the ML model significantly underpredicts the true error in its uncalibrated form. After calibration, the slope is now 1.02 and intercept is -0.01, showing a significant improvement to the predicted uncertainty values. The slopes and intercepts for additional recalibration tests, together with the best average recalibration parameters, are provided in **Table S4** of the **SI**. Overall, our present ML model built using an ensemble of neural networks provides reasonable uncertainty estimates when recalibrated. This analysis also points to the need for due diligence in evaluating uncertainty estimates for ML models, as the uncalibrated version of our ML model would have significantly underestimated the true uncertainty of predicted TTS values.

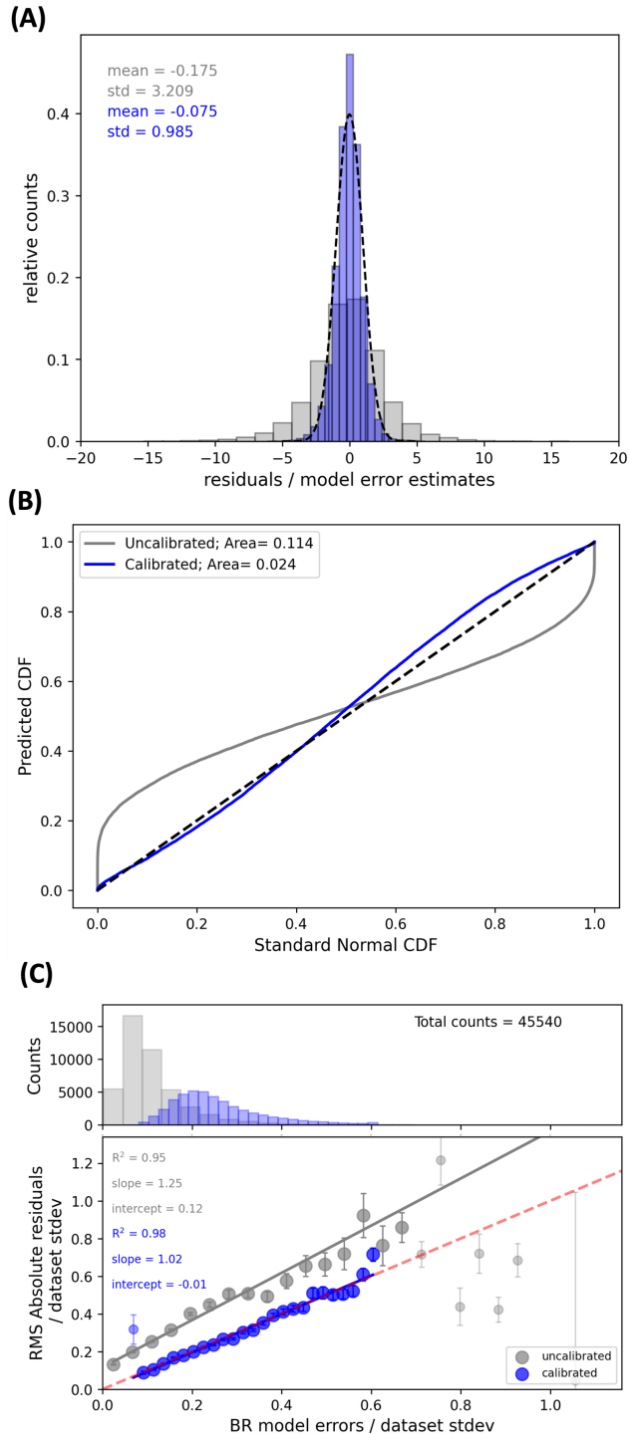

**Figure 7.** Plots assessing the quality of the uncertainty estimates of the ML model, before (grey) and after (blue) recalibration. (A) Distribution of normalized residuals (residuals divided by model error estimates. (B) Q-Q CDF plot of probability distributions of data from (A). The dashed line indicates a perfect unit-normal Gaussian distribution. (C) Residual vs. error (RvE) plot, where the red dashed line is the y=x line, and the histograms denote the distribution of numbers of data points comprised in each bin.

## 3.6 Examining physical trends of hardening with the ML model

In addition to providing accurate TTS predictions and well-calibrated uncertainties on those predictions, it is also important that our ML model demonstrates behavior consistent with known physics of RPV alloy embrittlement. Here, we examine the relative importance of the input features in predicting TTS using the SHapley Additive exPlanation (SHAP) approach[45] with a single neural network used in our ensemble, as shown in **Figure 8**. Briefly, SHAP is a mathematical method of explainable machine learning which uses game theoretic principles to assess the contributions of each feature in a model prediction, thus providing both ranked feature importances and trends of the target variable with each feature. From **Figure 8**, we see that the most important features for predicting TTS are the Cu content, the fluence (as log base 10), the Ni content, and the temperature. These top four features are consistent with the previous work of Ferreño et al.[25] and general physical understanding of the key factors controlling RPV steel embrittlement.[6] Furthermore, the trends in TTS with each feature are as expected from physical understanding. For example, the SHAP analysis shows that increased values of Cu content, fluence, and Ni content all tend to result in higher TTS, lower values of flux and temperature tend to lead to higher values of TTS, and Si and C content have little correlation with TTS. All of these trends are consistent with known trends from other studies and physical understanding, discussed more below.

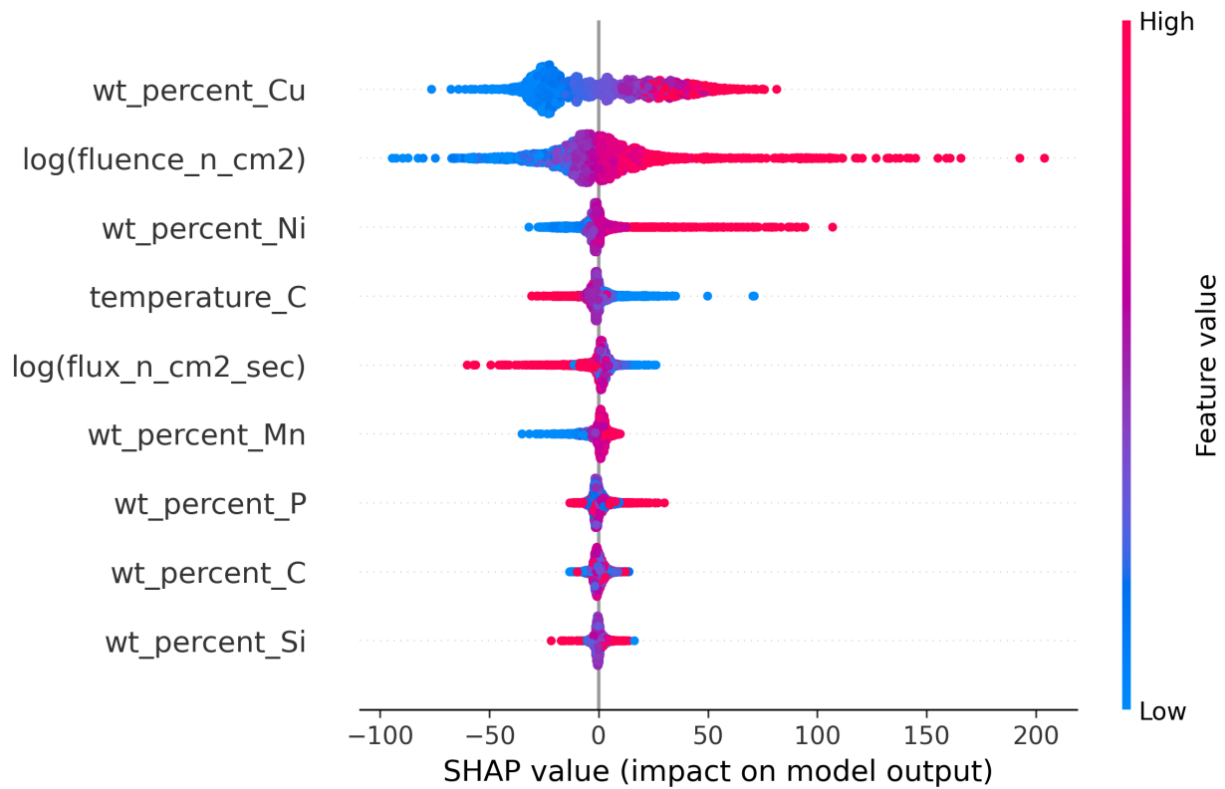

**Figure 8.** Summary of the SHapley Additive exPlanation (SHAP) results using a single neural network comprising our ensemble. The x-axis represents the average percentage contribution of each feature to the model predictions, representing a ranked list of feature importances.

Next, we examine the trends of our ML model predictions as a function of four key features for which embrittlement trends are well-known: Cu content (in weight percent), Ni content (in weight percent), fluence and flux. These results are presented in **Figure 9** in the form of cross-plots. We note that a version of these cross-plots with shifted data points overlaid is provided in **Figure S6** of the **SI**. Because we are plotting predicted TTS as a function of a single variable, we hold the other composition and temperature values fixed to correspond to median values in the full database. These median values correspond to a fictious alloy with temperature = 290 ˚C, wt% Cu = 0.12, wt% Ni = 0.69, wt% Mn = 1.4, wt% P = 0.009, wt% Si = 0.22, wt% C = 0.14. The flux and fluence values were chosen to correspond a flux of $3 \times 10^{10}$ n/cm²-s and fluences of $4 \times 10^{19}$ n/cm² (black dashed lines) and $1 \times 10^{20}$ n/cm² (green dashed lines). The exception to this is the plot of predicted TTS vs. fluence in **Figure 9C**, where the ML model was evaluated at different flux levels. In **Figure 9**, the ML predictions are shown as dashed lines, where the shaded regions are the corresponding one-standard deviation calibrated error bars. The TTS increases significantly with increasing Cu content (**Figure 9A**), which is consistent with physical understanding due to increased volume fractions of CMP dislocation glide obstacles, resulting in increased $\Delta\sigma_y$ and TTS. The leveling off at about 0.3 wt.% Cu is due to pre-precipitation during RPV heat treatment.[46,47] The further decrease in TTS at even higher Cu up to ≈0.8 wt% Cu is believed to be due to enhanced pre-precipitation driven by higher supersaturation of this insoluble element.[47,48] The increased TTS with Ni in **Figure 9B** is also due to an increase in the precipitate volume fractions, since precipitates typically contain more than 30% Ni. The predicted TTS with fluence in **Figure 9C** at lower flux (black dashed curve for typical RPV service conditions at ≈3×10$^{10}$ n/cm$^2$-s) flatten at lower fluence due to Cu depletion from the matrix, but subsequently increase significantly at fluences of greater than 10$^{19}$ n/cm$^2$; this is due to continuing precipitation of Ni, Mn and Si which is slower than that for Cu.[6] In **Figure 9C**, since we are plotting as a function of fluence, the ML predictions are given at the low flux of 3×10$^{10}$ n/cm$^2$-s, while the green dashed curve in **Figure 9C** is for a very high flux conditions of 1×10$^{14}$ n/cm$^2$-s and shows a generally lower TTS up to high fluence. It is well-established that high flux delays precipitation because vacancy and interstitial recombination lead to less efficient radiation-enhanced diffusion.[4,6,49,50] However, it is also well known that the effect of flux decreases with increasing fluence in a way that depends on the steel composition.[6,28] At very high fluence greater than 10$^{20}$ n/cm$^2$ there is a crossover with higher TTS at higher flux. However, such a crossover is not observed at lower fluxes closer to service conditions, and physical considerations suggest that it is likely not reliable.[6,28] The predicted TTS shown in **Figure 9D** shows that the TTS continuously decreases with flux at an intermediate fluence of 4x10$^{19}$ n/cm$^2$. In contrast, at a higher fluence of 10$^{20}$ n/cm$^2$, the predicted TTS varies non-monotonically as a function of flux; however, the overall variation in TTS is modest, and falls well within the uncertainty band, ranging from a low to high value of 98 to 110 °C, respectively, over the entire range of flux.

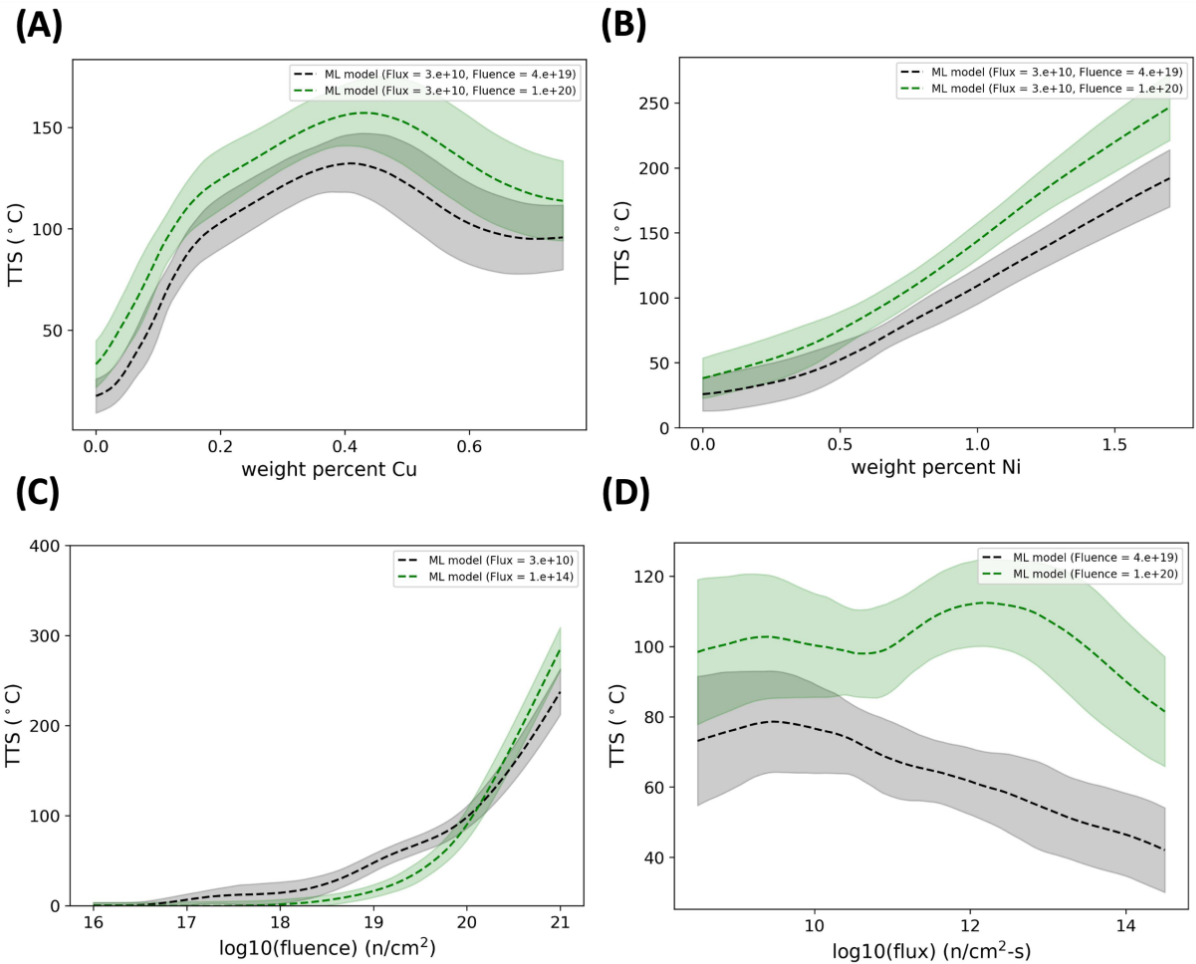

**Figure 9.** Cross-plots showing trends in ML model predictions as a function of key variables: (A) TTS as a function of wt% Cu, (B) TTS as a function of wt% Ni, (C) TTS as a function of fluence, and (D) TTS as a function of flux. All of these plots assume composition and temperature values (where appropriate) for a median alloy in the database (temperature = 290 ˚C, wt% Cu = 0.12, wt% Ni = 0.69, wt% Mn = 1.4, wt% P = 0.009, wt% Si = 0.22, wt% C = 0.14). The dashed lines in each plot are the ML model predictions under the flux and fluence conditions indicated in the legend, and the shaded region is the calibrated one-standard deviation error bar of the ML model.

In **Figure S7** of the **SI**, we provide the flux cross plots of all 328 US plant surveillance alloys at various fluence levels. Overall, the flux behavior is essentially flat over the flux range of interest for life extension conditions (i.e., log flux of 10-11 n/cm²-s) for all fluence levels. In addition, the fluence cross plots shown in **Figure 10** for six principal alloys shows physically reasonable flux effects up to life extension conditions of fluence of $10^{20}$ n/cm², except in low Cu (LG). There is a significant literature on flux effects.[4,49–60] However, this research has led to different conclusions, ranging from strong to no flux effects. It is now clear that the controversies related to flux effects were mainly the result of complex interactions between all significant compositional (Cu, Ni, Mn…) and irradiation (temperature, flux, fluence) variables (ML features) which mediate TTS. Most significantly, there is now strong empirical evidence that flux effects are significant at low fluence, particularly in sensitive Cu bearing steels, but they decrease, or disappear, at high fluence.[6,28,60] For example, this conclusion was specifically supported both in the extensive hand analysis of test reactor data,[6] and in the ML study by Liu et al.[28]

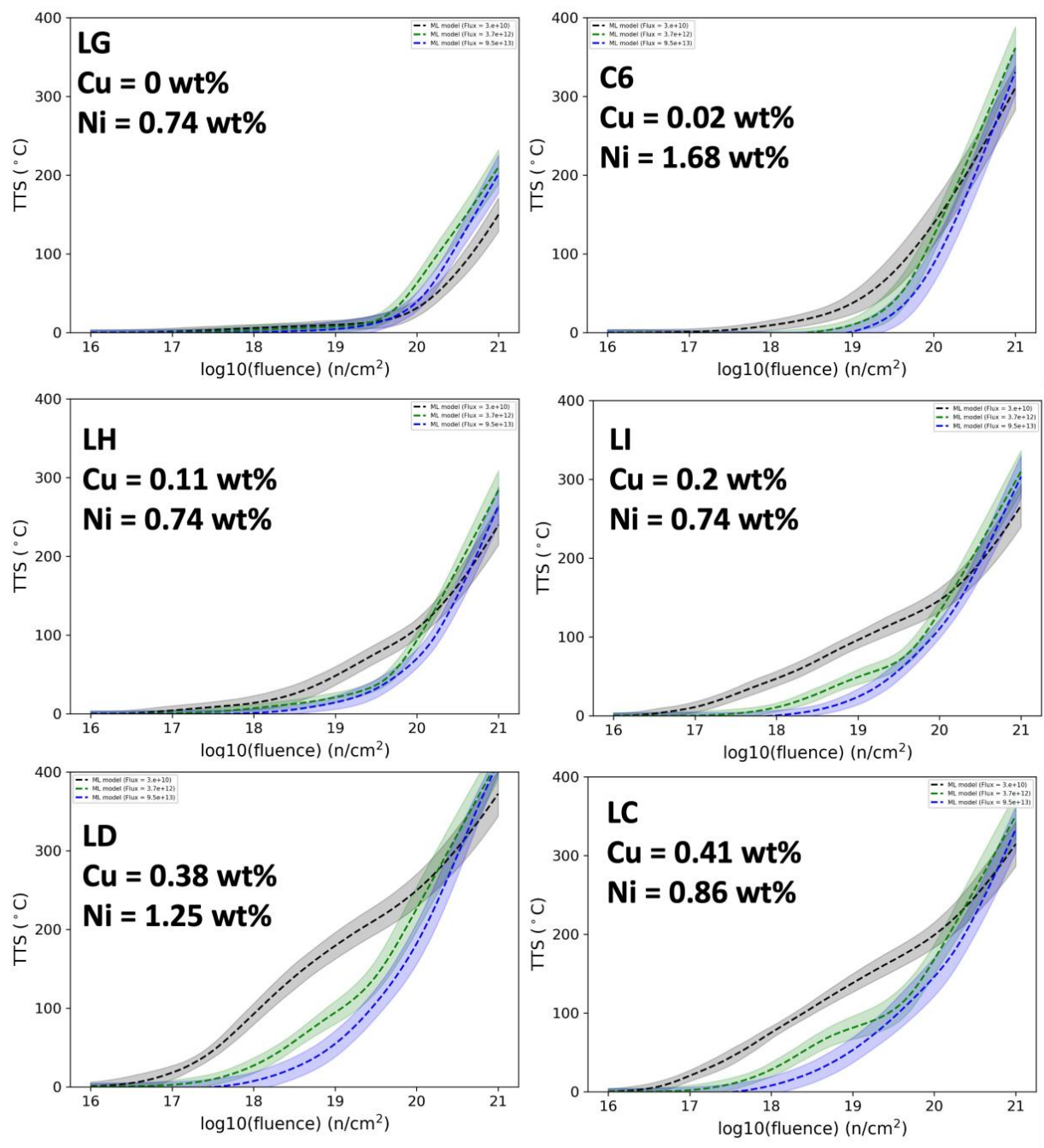

**Figure 10.** Fluence cross-plots for six principal alloys with increasing Cu content. The dashed lines in each plot are the ML model predictions under the flux conditions indicated in the legend, and the shaded region is the calibrated one-standard deviation error bar of the ML model. Three flux values were selected: a low flux of $3\times10^{10}$ n/cm²-s (black) representative of typical reactor operation, and high flux values of $3.68\times10^{12}$ (green) and $9.5\times10^{13}$ n/cm²-s (blue) representative of ATR2 and BR2 flux conditions, respectively.

Specifically, the ML predictions in this study (see **Figure 10**) clearly show increasing flux delays, thus decreases TTS at low to intermediate fluence, except in low Cu steels. However, at higher fluence, ML predicts a cross-over in Cu bearing steels, where higher flux leads to higher TTS. As shown in **Figure 10**, the predicted crossover fluence is very low in low Cu (< ≈ 0.075%) and low to medium Ni (≤ 0.75%) steels. At higher Cu (> ≈ 0.1%) the crossover fluence increases with Ni. Although details differ, a qualitatively similar trend was observed in the test reactor ML study of $\Delta\sigma_y$ in Liu et al.[28] Note the crossover fluence also depends on the specific values of the high and low fluxes. Thus, while the fluence at a specified $\Delta\sigma_y$ was well-predicted by the earlier ML model in domains with flux-fluence data, significant extrapolations of the simpler curve shapes were taken as being unreliable. Based on these considerations, the previous ML study from Liu et al. assumed there is no effect of flux beyond the crossover fluence.[28]

The most relevant high and low fluxes are the ATR2 values of $3.68 \times 10^{12}$ n/cm²-s, and a typical PWR vessel flux of ≈3-4x10^10 n/cm²-s, which reaches a fluence of $10^{20}$ n/cm² in about 80-100 years. In the current study the crossover fluence in Cu bearing steels is near, or more generally above, $10^{20}$ n/cm²-s (see **Figure 10**). In contrast, the crossover is at very low fluence, and the TTS decreases with decreasing flux, in low Cu steels. It is beyond the scope of the current paper to explore the physics that might support this observation in low Cu steels, such as the role of so-called unstable matrix defects, which have opposing effects of increasing recombination, while adding a small increment of hardening.[51–54] Overall, additional study to better assess the applicability domain of our model and more deeply analyze the flux effects of specific alloys is warranted. In summary, our ML model TTS predictions are broadly consistent with both previously observed and key physical mechanisms, which mediate embrittlement of RPV steels.

### 3.7    TTS predictions for life extension

Having established that our ML model provides accurate embrittlement predictions with well-calibrated uncertainties, and that the embrittlement predictions follow known physical trends, we next predict TTS values and their uncertainties encompassing approximate LWR life extension conditions. **Figure 11A** contains a heatmap grid of predicted TTS values as a function of flux and fluence for the same fictitious "median" alloy used for the TTS predictions in **Figure 9**.

In **Figure 11A**, the colors denote the magnitude of TTS value (blue is lower, red is higher), the large numbers in the first row of each box are the predicted TTS in ˚C, and the smaller number underneath is the corresponding calibrated one standard deviation uncertainty estimate. The values corresponding to a fluence of $10^{20}$ n/cm$^2$ are outlined with a black box. From these predictions, we see that a hypothetical median alloy has a TTS of about 100-101 ˚C +/- 13-15 ˚C. **Figure 11B** plots the predicted TTS values for 328 unique alloys in the US reactor fleet in the PLOTTER database. We have provided the corresponding plot of predicted TTS values for all 723 alloys in **Figure S8** of the **SI**, and a parity plot comparing our ML predictions with the E900 model for the subset of 328 reactors in the US fleet is shown in **Figure S9** of the **SI**. Values for our ML model (blue), E900 (green) and EONY (purple) models are provided, sorted from low to high ML TTS predictions. The shaded blue regions are the calibrated one standard deviation uncertainty estimates. From **Figure 11B**, it is evident the behavior of E900 and EONY models is qualitatively similar to our ML model in predicting TTS of each alloy, with EONY showing a slight tendency to underestimate TTS relative to the ML model. Given the prevalence of the E900 model in predicting RPV embrittlement, we focus our discussion to comparing the behavior of our ML model predictions with E900. In more than half of the alloys (179 out of 328 or 54.5%), the TTS difference between E900 and our ML model predictions are between in the range -8 to +8 ˚C, which is roughly the same, or less than, the experimental uncertainty of the individual TTS measurement.[25] In addition, 261 of the 328 (79.6%) alloys have E900 TTS values that are within the one-sigma calibrated error bar of the ML model uncertainty, and 314 of them (96.6%) are within two-sigma. When examining the top 10 alloys with highest predicted TTS values for each model, 5 of the 10 same alloys are present in the top 10 for ML and E900 models, and 6 of the 10 same alloys are present for the EONY and E900 models, suggesting consistent ranking of alloys with high TTS across different models.

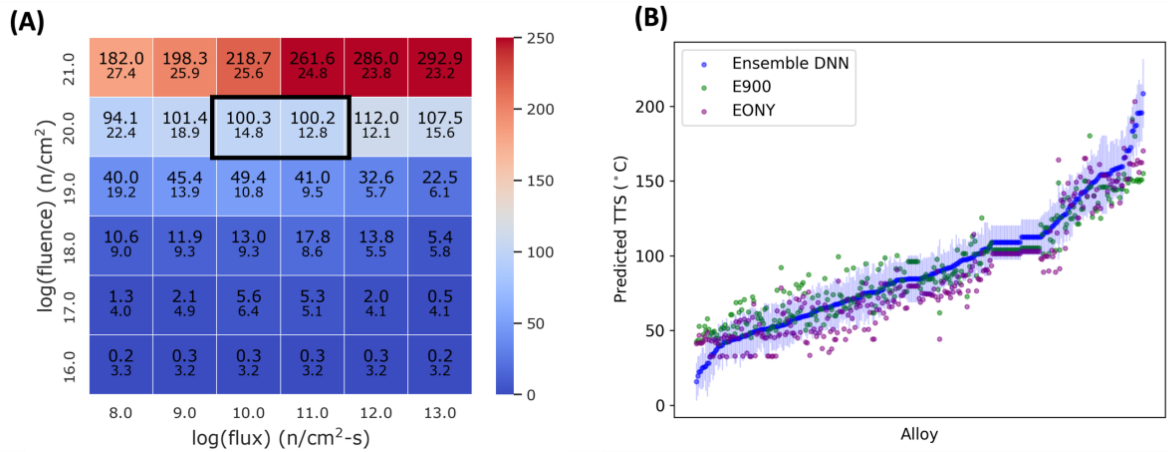

**Figure 11.** (A) Heatmap showing the predicted TTS values as a function of flux and fluence for the same median alloy as considered in **Figure 9**. The large numbers denote the ML-predicted TTS values, and the smaller numbers underneath are the calibrated uncertainty values (one standard deviation). The values corresponding to approximate 100-year life extension conditions are shown in the black box. (B) Values of predicted TTS for each unique alloy in the US reactor fleet (328 in total) in the PLOTTER database. The values are sorted by the ML-predicted values (blue data, where the shaded blue region is the calibrated uncertainty value range (one standard deviation)). The green and purple points denote the corresponding TTS predictions from the E900 and EONY models, respectively.

## 4.    Conclusion:

Despite the successes of ML models in the literature and from the present work, models that are more accurate and apply to larger application domains remain desirable. Here, we offer some perspective ideas for how ML models to predict embrittlement might be improved in future studies. One is to include additional relevant information related to material testing to better include potentially relevant variables not in the current model. A concrete recent example from Erickson and Kirk[33] added hundreds of unirradiated yield strengths and unirradiated ductile-to-brittle transition temperatures to the PLOTTER database. Erickson and Kirk modified the ASTM E900 model to have additional terms including the effect of unirradiated yield strengths, and they found an RMSE reduction from 13.3 °C to 13 °C, roughly a 3% error reduction. Using these initial, unirradiated mechanical property values as features may provide a means to further reduce the ML model error. For example, a 5-fold CV fit of our ML model to the subset of 249 PLOTTER values which contain unirradiated yield strength and unirradiated T41J values and comparing the fits obtained with and without these unirradiated values as features in the model results in RMSE values of 11.0 °C and 10.5 °C, respectively, a 4.6% RMSE reduction. However, only a small fraction of the PLOTTER-22 database contains information on the unirradiated properties of RPV steels, and fleshing out the database with additional unirradiated property values could be beneficial for improved model fits. Another potential area for model refinement is to develop databases which contain both TTS and $\Delta\sigma_Y$ values. In this work, we used a simple 2-term quadratic fit to ≈300 experimental measurements of TTS and $\Delta\sigma_Y$. However, this conversion is alloy- and irradiation condition-dependent, so a major opportunity exists to develop more quantitative models, including variables that mediate the TTS - $\Delta\sigma_Y$ relation.

**Acknowledgements:** This work was funding by the US Department of Energy (DOE) Nuclear Energy University Program (NEUP) under award number DE-NE0009143. The authors gratefully acknowledge Mark Kirk, who provided us the PLOTTER-22 database and many helpful discussions about the PLOTTER-22 data and the E900 model, and for providing an early review and feedback on a draft of this paper.

**Ethical statement:** The authors have no conflicts of interest to declare.

**Data and Code Availability:** The raw data required to reproduce these findings are available upon request. The UCSB dataset is available on Figshare: https://doi.org/10.6084/m9.figshare.23304227. The PLOTTER dataset is available online: https://www.astm.org/adje090015-ea.html. The final model fit to the entire database is available on Github: https://github.com/uw-cmg/RPV_model. The model can be run on Google Colab via the link in the Github repository. This model provides predicted TTS values and the calibrated one standard deviation error bar.

**Supporting Information for:**

**Predictions and Uncertainty Estimates of Reactor Pressure Vessel Steel Embrittlement using Machine Learning**

**Authors:** Ryan Jacobs[1], Takuya Yamamoto[2], Dane Morgan[1], G. Robert Odette[2]

[1] Department of Materials Science and Engineering, University of Wisconsin-Madison, Madison, WI, USA

[2] Mechanical Engineering Department, University of California Santa Barbara, Santa Barbara, CA, USA.

**Table S4.** Summary of key statistics of the RPV embrittlement databases used in this work.

| | | ATR2 | BR2 | IVAR | PLOTTER | RADAMO | PLOTTER-15 MTR | Full database |
|---|---|---|---|---|---|---|---|---|
| **Number of points** | | 50 | 36 | 1464 | 2048 | 342 | 595 | 4535 |
| **Transition temperature shift (˚C)** | Min | 55.89 | 0.0 | -7.67 | -28.9 | -7.67 | -14.0 | -28.93 |
| | Max | 370.83 | 329.27 | 286.96 | 253.0 | 276.98 | 306.78 | 370.83 |
| | Mean | 158.35 | 96.85 | 48.13 | 42.1 | 61.10 | 78.03 | 51.84 |
| | Stdev | 59.70 | 70.56 | 43.86 | 37.1 | 57.92 | 45.11 | 46.15 |
| **Temperature (˚C)** | Min | 290.0 | 290.0 | 270.0 | 255.0 | 265.0 | 250.0 | 250.0 |
| | Max | 290.0 | 300.0 | 310.0 | 303.9 | 300.0 | 315.0 | 315.0 |
| | Mean | 290.0 | 296.7 | 290.0 | 284.8 | 293.86 | 288.33 | 287.79 |
| | Stdev | 0.0 | 4.7 | 12.2 | 6.9 | 13.31 | 5.56 | 9.81 |
| **Cu (wt %)** | Min | 0.0 | 0.0 | 0.0 | 0.01 | 0.05 | 0.002 | 0.0 |
| | Max | 0.43 | 0.41 | 0.86 | 0.41 | 0.31 | 0.43 | 0.86 |
| | Mean | 0.19 | 0.19 | 0.23 | 0.11 | 0.15 | 0.14 | 0.16 |
| | Stdev | 0.14 | 0.16 | 0.19 | 0.08 | 0.09 | 0.10 | 0.14 |
| **Ni (wt %)** | Min | 0.07 | 0.74 | 0.00 | 0.04 | 0.6 | 0.07 | 0.00 |
| | Max | 1.70 | 1.68 | 1.71 | 1.70 | 1.01 | 1.85 | 1.85 |
| | Mean | 0.81 | 1.00 | 0.77 | 0.62 | 0.73 | 0.70 | 0.69 |
| | Stdev | 0.38 | 0.35 | 0.41 | 0.25 | 0.13 | 0.25 | 0.32 |
| **Mn (wt %)** | Min | 0.26 | 1.37 | 0.01 | 0.58 | 1.29 | 0.0 | 0.0 |
| | Max | 1.70 | 1.50 | 1.70 | 1.98 | 1.6 | 1.76 | 1.98 |
| | Mean | 1.37 | 1.41 | 1.43 | 1.33 | 1.46 | 1.18 | 1.35 |
| | Stdev | 0.32 | 0.05 | 0.33 | 0.25 | 0.11 | 0.47 | 0.32 |
| **P (wt %)** | Min | 0.002 | 0.005 | 0.002 | 0.002 | 0.005 | 0.003 | 0.002 |
| | Max | 0.050 | 0.007 | 0.050 | 0.024 | 0.021 | 0.028 | 0.050 |
| | Mean | 0.010 | 0.005 | 0.009 | 0.011 | 0.010 | 0.012 | 0.010 |
| | Stdev | 0.009 | 0.001 | 0.010 | 0.004 | 0.005 | 0.005 | 0.007 |
| **Si (wt %)** | Min | 0.13 | 0.17 | 0.01 | 0.00 | 0.04 | 0.00 | 0.00 |
| | Max | 0.63 | 0.24 | 0.63 | 1.00 | 0.45 | 0.27 | 1.00 |
| | Mean | 0.29 | 0.22 | 0.25 | 0.23 | 0.29 | 0.004 | 0.21 |
| | Stdev | 0.14 | 0.02 | 0.14 | 0.14 | 0.10 | 0.027 | 0.15 |
| **C (wt %)** | Min | 0.06 | 0.14 | 0.00 | 0.00 | 0.05 | 0.00 | 0.00 |
| | Max | 0.29 | 0.19 | 0.34 | 0.30 | 0.26 | 0.00 | 0.34 |
| | Mean | 0.15 | 0.16 | 0.14 | 0.14 | 0.15 | 0.00 | 0.12 |
| | Stdev | 0.05 | 0.02 | 0.05 | 0.08 | 0.06 | 0.00 | 0.08 |
| **Log Fluence (n/cm²)** | Min | 20.14 | 19.23 | 16.78 | 15.97 | 18.13 | 18.00 | 15.97 |
| | Max | 20.14 | 20.34 | 19.52 | 20.33 | 20.24 | 20.23 | 20.34 |
| | Mean | 20.14 | 19.83 | 18.53 | 18.95 | 19.23 | 19.41 | 18.91 |
| | Stdev | 0.00 | 0.37 | 0.58 | 0.69 | 0.62 | 0.38 | 0.70 |
| **Log Flux (n/cm²-s)** | Min | 12.57 | 13.32 | 10.85 | 8.36 | 12.30 | 11.19 | 8.36 |
| | Max | 12.57 | 13.99 | 12.00 | 12.73 | 13.98 | 14.00 | 14.00 |
| | Mean | 12.57 | 13.76 | 11.52 | 10.74 | 13.38 | 12.83 | 11.51 |
| | Stdev | 0.00 | 0.31 | 0.40 | 0.65 | 0.42 | 0.48 | 1.05 |

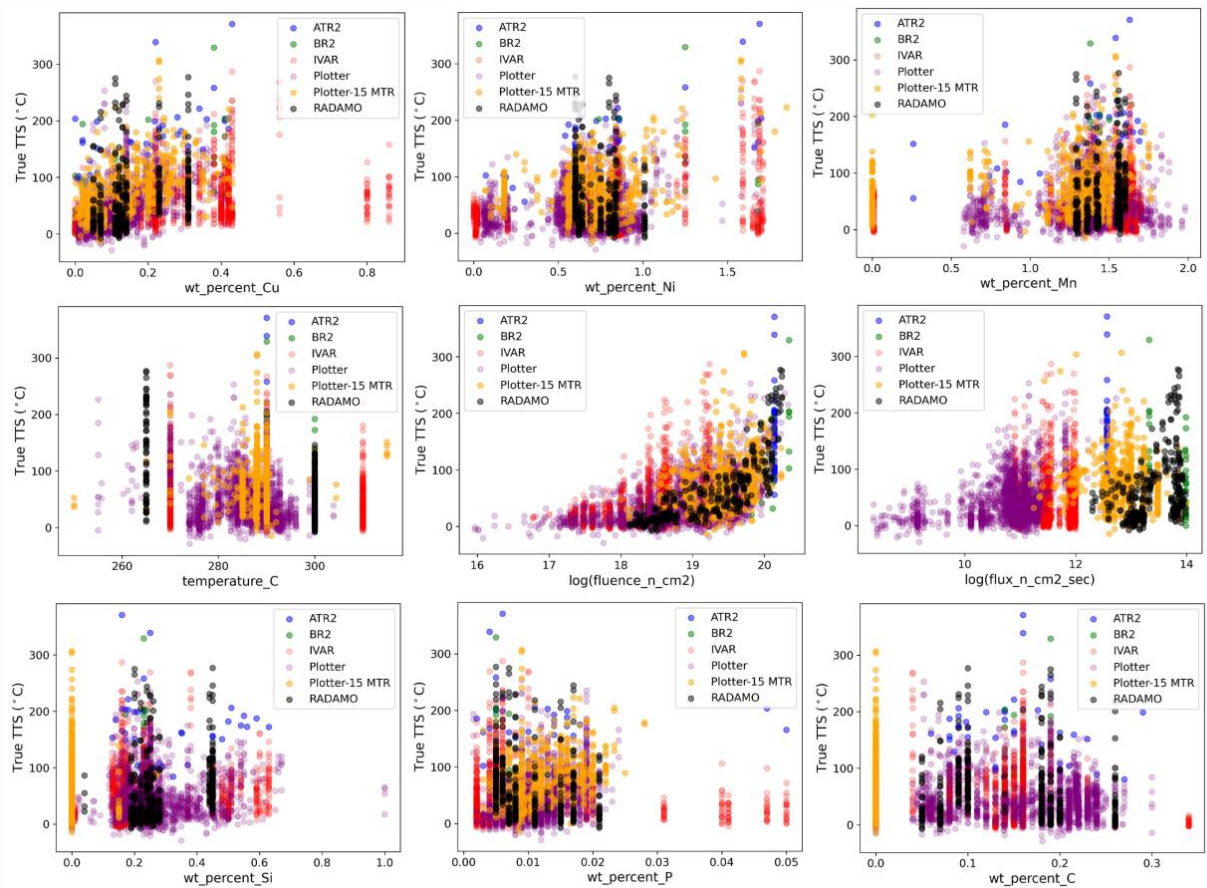

**Figure S1**. Scatter plots of the true TTS values as a function of each feature. Each figure panel denotes a different feature, as indicated by the x-axis title. In each figure, the blue, green, red, purple, orange and black points denote data corresponding to the ATR2, BR2, IVAR, Plotter, Plotter-15 MTR and RADAMO data groups, respectively.

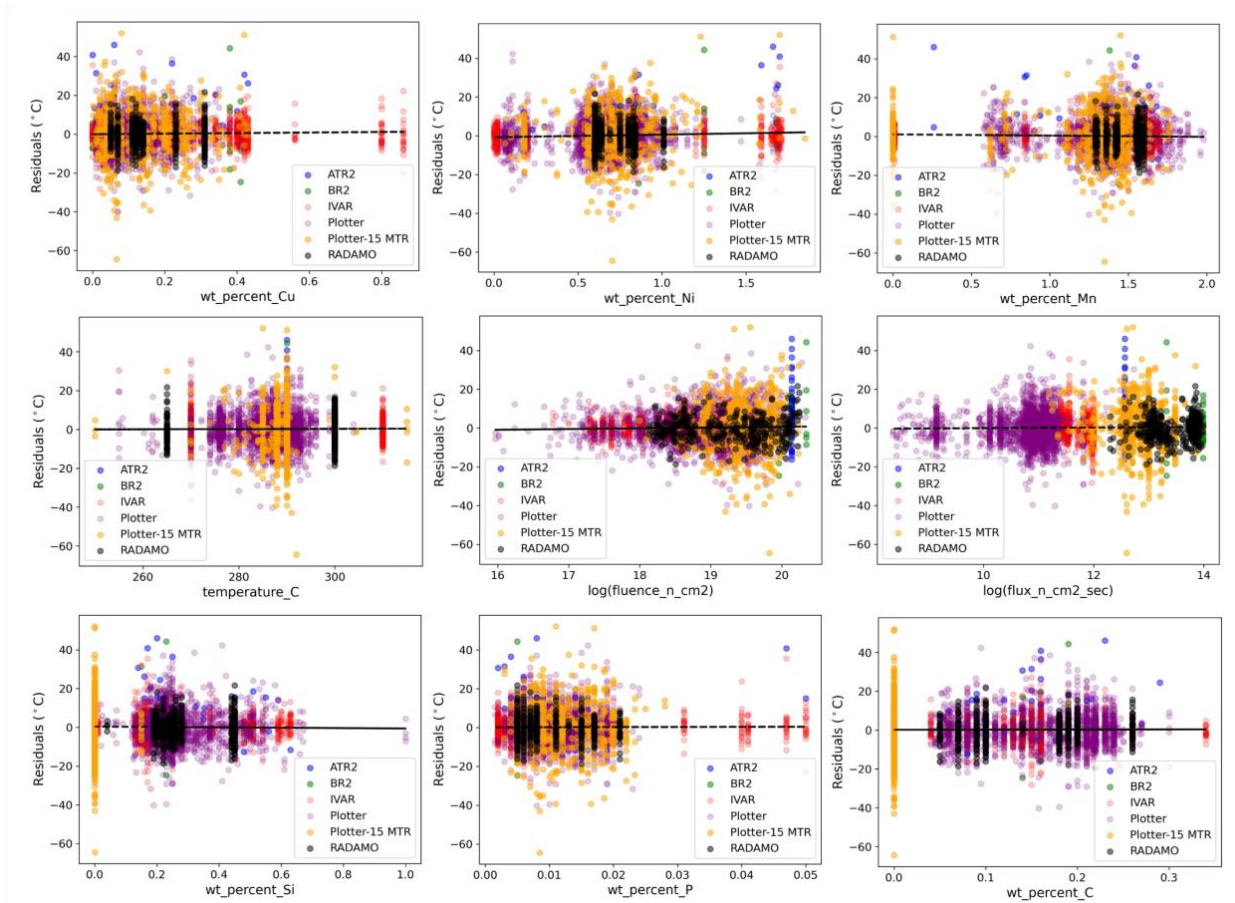

**Figure S2**. Scatter plots of the residual TTS values as a function of each feature (residuals = true minus ML-predicted TTS values) for the full fit to the combined database. Each figure panel denotes a different feature, as indicated by the x-axis title. In each figure, the blue, green, red, purple, orange and black points denote data corresponding to the ATR2, BR2, IVAR, Plotter, Plotter-15 MTR and RADAMO data groups, respectively. A summary of linear fit statistics is provided in **Table S2**.

**Table S2**. Summary of linear fit statistics for the residual vs. feature plots shown in **Figure S3**. To aid in interpreting the impact of the slopes, the change in residual TTS over the dynamic range of the data is also provided, and is small in all cases, indicating negligible trend of residuals vs. each feature.

| Feature | Slope | Intercept | TTS change (℃) |
|---|---|---|---|
| wt_percent_Cu | 1.452 | -0.016 | 1.23 |
| wt_percent_Ni | 1.358 | -0.731 | 1.78 |
| wt_percent_Mn | -0.642 | 1.081 | -0.19 |
| temperature_C | 0.006 | -1.421 | -1.03 |
| log(fluence_n_cm2) | 0.379 | -6.954 | -5.30 |
| log(flux_n_cm2_sec) | 0.172 | -1.766 | -0.80 |
| wt_percent_Si | -0.99 | 0.422 | -0.57 |
| wt_percent_P | 5.002 | 0.159 | 0.40 |
| wt_percent_C | 0.591 | 0.138 | 0.34 |

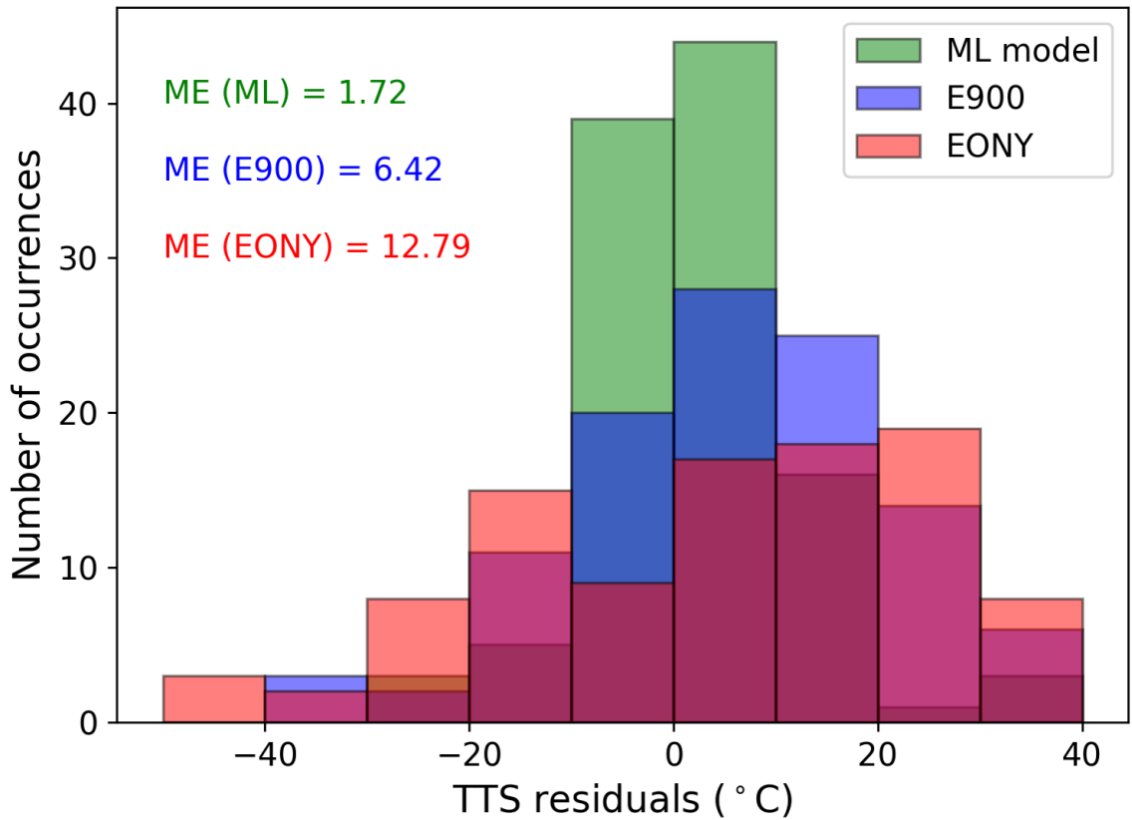

**Figure S3**. Histograms of residuals for the high fluence subset of PLOTTER data using our full fit ML model (green), the E900 model (blue), and the EONY model (red). The mean error (ME) values are in ˚C. The standard deviations of the residuals are 10.41, 16.56 and 38.46 ˚C for the ML model, E900 and EONY models, respectively.

**Table S3.** Summary of MAE values of various ML models from our work and the literature

| Study | Dataset | Model type | MAE (˚C) | Notes |
|---|---|---|---|---|
| This work | PLOTTER-22 | Ensemble NN | 6.4 | Full fit |
| This work | PLOTTER-22 | Ensemble NN | 8.8 | 5-fold CV (25 splits) |
| Ferreño et al.[25] | PLOTTER-15 | Gradient boosting | 8.37 | Single test split |
| Mathew et al.[22] | NUREG (old subset of PLOTTER-15) | Bayesian NN | 30.7 MPa × 0.6 = 18.4 | Single test split |
| This work | IVAR | Ensemble NN | 3.8 | Full fit |
| This work | IVAR | Ensemble NN | 5.4 | 5-fold CV (25 splits) |
| Mathew et al.[22] | IVAR | Bayesian NN | 16.1 MPa × 0.6 = 9.66 | Single test split |
| Liu et al.[24] | IVAR+ | Kernel ridge regression | 14.8 × 0.6 = 8.88 | 5-fold CV |

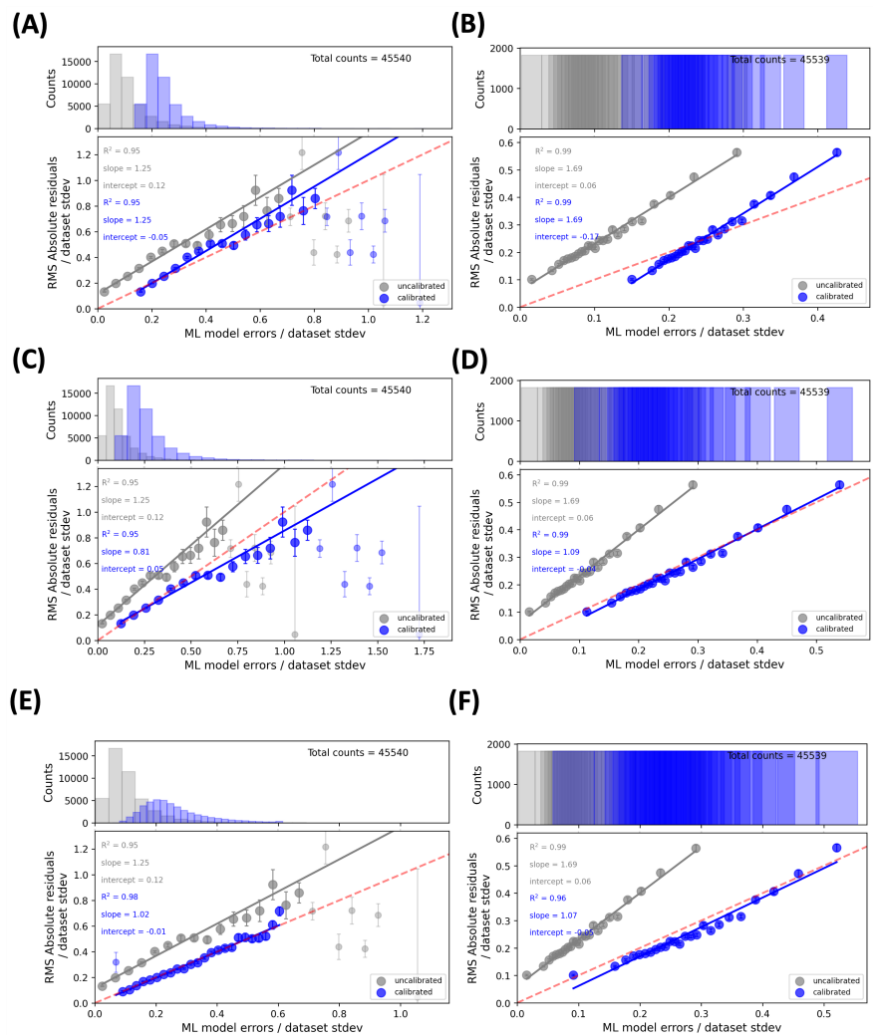

**Figure S4.** RvE plots corresponding to constant rescaling ((A) and (B)), linear rescaling ((C) and (D)), and quadratic rescaling ((E) and (F)). All plots are made with 25 total bins. Plots (B), (D) and (F) are made such that an equal number of points resides in each bin.

**Table S4.** Summary of various error bar recalibration schemes. The values are averages +/- standard deviations across all 50 nested cross validation splits. In the second column, the optimized function values are half the negative-log likelihood (NLL) values divided by the number of residuals in the optimization, i.e., 0.5×NLL/(number of residuals). For the recalibrated RvE plot $R^2$, slope and intercept values, the first number is the value with non-constant bin sizes, and the second value (in parentheses) is the value with constant bin sizes.

| Recalibration scheme | 0.5×NLL/(number of residuals) | Recalibration parameters | $R^2$ | slope | intercept |
|---|---|---|---|---|---|
| Constant | 3.827 +/- 0.014 | 6.211 +/- 0.0152 | 0.95 (0.99) | 1.25 (1.69) | -0.05 (-0.17) |
| Linear | 3.814 +/- 0.014 | 1.548 +/- 0.075; 4.077 +/- 0.303 | 0.95 (0.99) | 0.81 (1.09) | 0.05 (-0.04) |
| Quadratic | 3.817 +/- 0.018 | -0.041 +/- 0.025; 2.041 +/- 0.344; 3.124 +/- 0.717 | 0.98 (0.96) | 1.02 (1.07) | -0.01 (-0.05) |

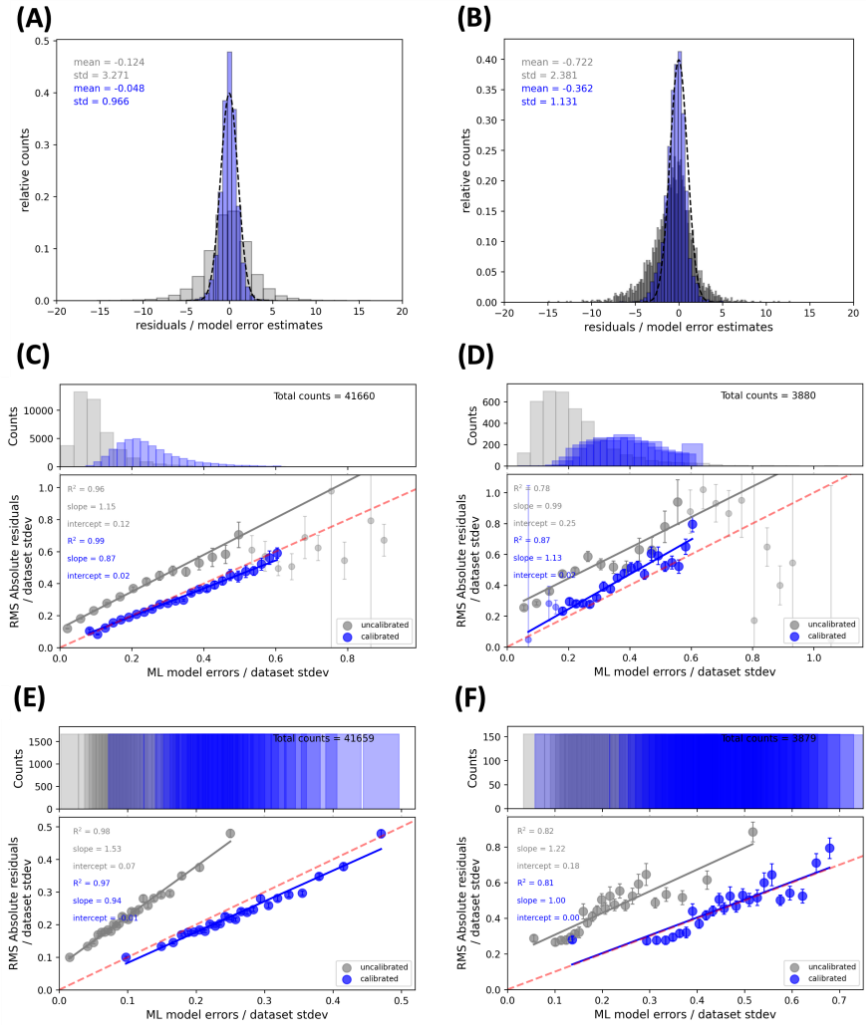

**Figure S5.** Normalized residual and RvE plots with quadratic recalibration for low fluence ((A), (C), (E)) and high fluence ((B), (D), (F)) subsets of data. All plots are made with 25 total bins. Plots (E) and (F) are made such that an equal number of points resides in each bin.

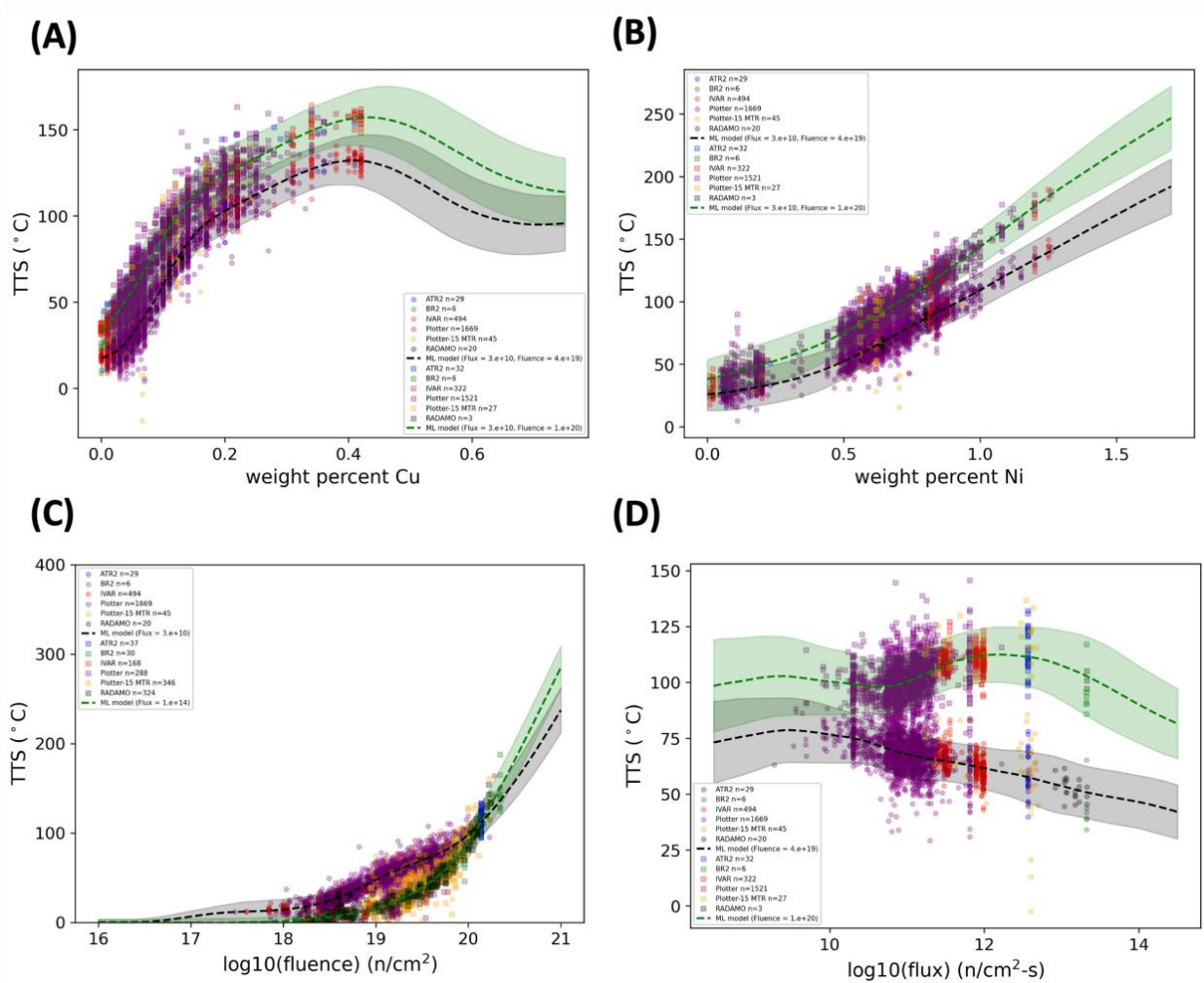

**Figure S6.** Cross-plots showing trends in ML model predictions as a function of key variables: (A) TTS as a function of wt% Cu, (B) TTS as a function of wt% Ni, (C) TTS as a function of fluence, and (D) TTS as a function of flux. All of these plots assume composition and temperature values (where appropriate) for a median alloy in the database (temperature = 290 °C, wt% Cu = 0.12, wt% Ni = 0.69, wt% Mn = 1.4, wt% P = 0.009, wt% Si = 0.22, wt% C = 0.14). The dashed lines in each plot are the ML model predictions under the flux and fluence conditions indicated in the legend, and the shaded region is the calibrated one-standard deviation error bar of the ML model. The data points are nearby (to the median alloy and flux/fluence feature values) experimental data points which have been shifted by their respective ML predictions. These data points were selected by using a sum of squares value of 10 in reduced units, where each feature was normalized to have mean zero and standard deviation of one.

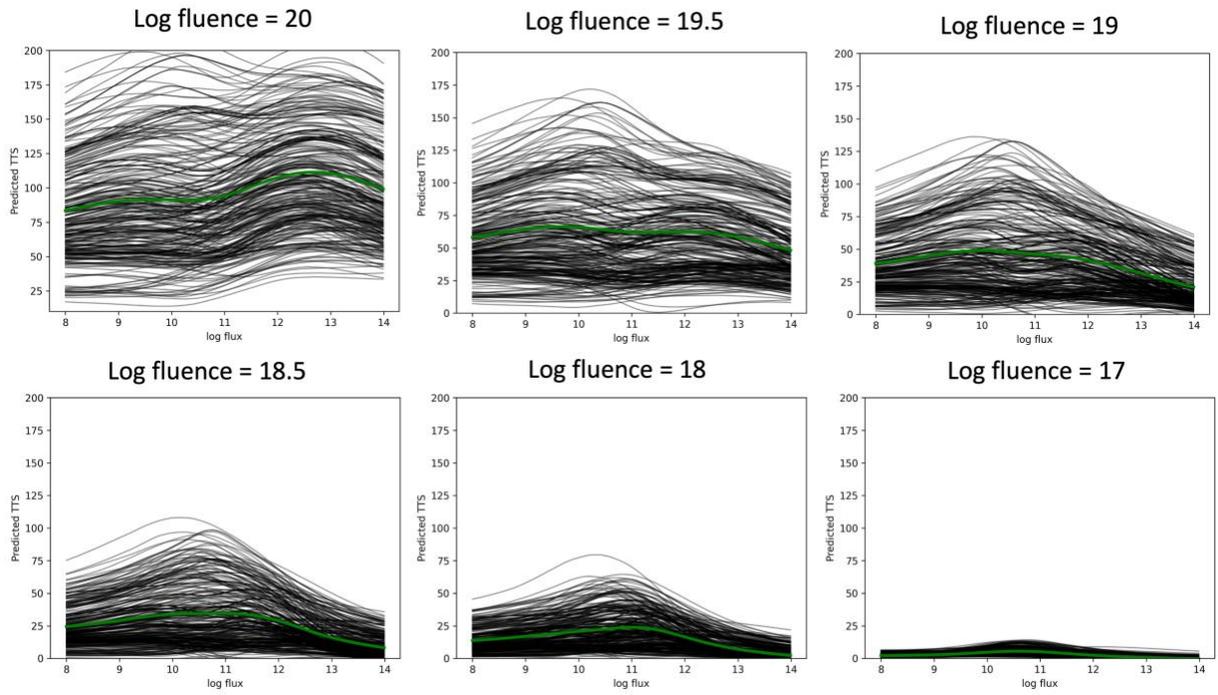

**Figure S7.** Flux cross plots for all 328 US reactor fleet alloys at various fluence levels. The black lines in each plot are the flux cross plot of an individual alloy, and the green lines are the average flux behavior.

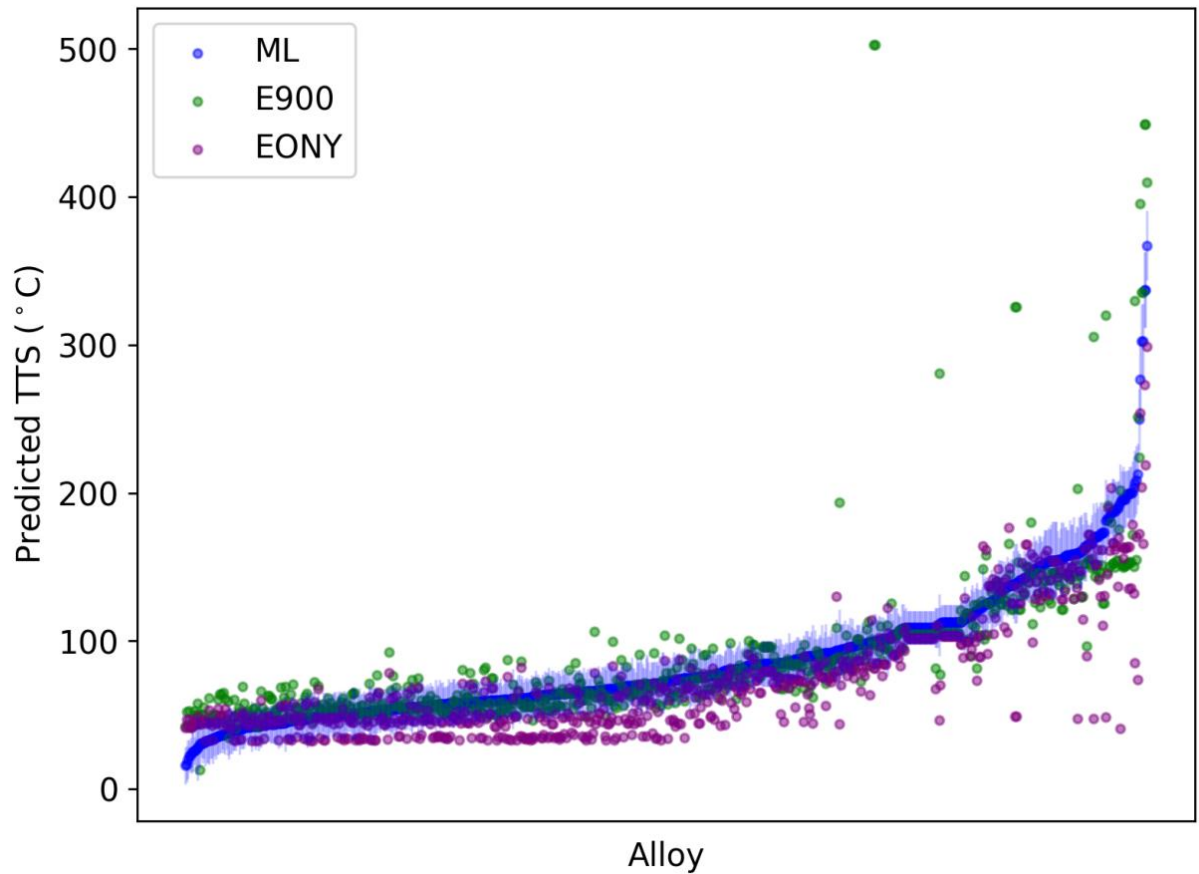

**Figure S8.** Values of predicted TTS for each unique alloy (723 in total) in the combined UCSB+PLOTTER database. The values are sorted by the ML-predicted values (blue data, where the shaded blue region is the calibrated uncertainty value range (one standard deviation)). The green and purple points denote the corresponding TTS predictions from the E900 and EONY models, respectively.

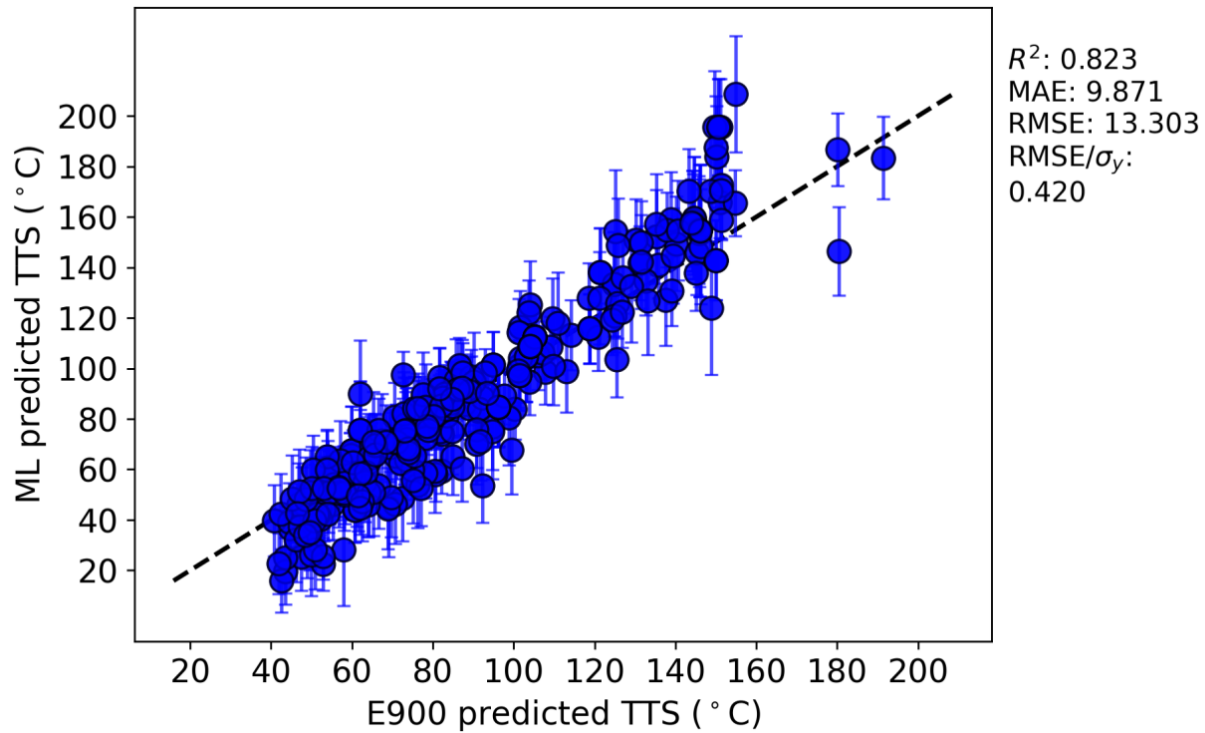

**Figure S9.** Parity plot of TTS predictions from our ML model vs. E900 model for the subset of 328 alloys in the US reactor fleet. The error bars are the calibrated one-standard deviation ML error bars.

## References

(1)     Organization, W. N. *Nuclear Power in the European Union*. https://world-nuclear.org/information-library/country-profiles/others/european-union.aspx (accessed 2022-11-02).

(2)     Organization, W. N. *Nuclear Power in the USA*. https://world-nuclear.org/information-library/country-profiles/countries-t-z/usa-nuclear-power.aspx (accessed 2022-11-02).

(3)     IPCC. *Climate Change 2022: Impacts, Adaptation and Vulnerability*; 2022.

(4)     Eason, E. D.; Odette, G. R.; Nanstad, R. K.; Yamamoto, T. A Physically-Based Correlation of Irradiation-Induced Transition Temperature Shifts for RPV Steels. *Journal of Nuclear Materials* **2013**, *433* (1–3), 240–254. https://doi.org/10.1016/j.jnucmat.2012.09.012.

(5)     Odette, G. R.; Lucas, G. E. Embrittlement of Nuclear Reactor Pressure Vessels. *JOM* **2001**, *53* (7), 18–22. https://doi.org/10.1007/s11837-001-0081-0.

(6)     Odette, G. R.; Yamamoto, T.; Williams, T. J.; Nanstad, R. K.; English, C. A. On the History and Status of Reactor Pressure Vessel Steel Ductile to Brittle Transition Temperature Shift Prediction Models. *Journal of Nuclear Materials* **2019**, *526*, 151863. https://doi.org/10.1016/j.jnucmat.2019.151863.

(7)     *ASTM International: Standard Practice for Design of Surveillance Programs for Light-Water Moderated Nuclear Power Reactor Vessels*; 2015. https://doi.org/10.1520/E0185-15.Copyright.

(8) Hashimoto, Y.; Nomoto, A.; Kirk, M.; Nishida, K. Development of New Embrittlement Trend Curve Based on Japanese Surveillance and Atom Probe Tomography Data. *Journal of Nuclear Materials* **2021**, *553*. https://doi.org/10.1016/j.jnucmat.2021.153007.

(9) *ASTM International (2015) Standard Guide for Predicting Radiation-Induced Transition Temperature Shift in Reactor Vessel Materials, E900-15*; 2015. https://doi.org/10.1520/E0900-15.

(10) Morgan, D.; Jacobs, R. Opportunities and Challenges for Machine Learning in Materials Science. *Annu Rev Mater Res* **2020**, *50*, 71–103. https://doi.org/10.1146/annurev-matsci-070218-010015.

(11) Mueller, T.; Kusne, A. G.; Ramprasad, R. Machine Learning in Materials Science: Recent Progress and Emerging Applications. In *Reviews in Computational Chemistry*; John Wiley & Sons, 2016; pp 186–273.

(12) Dimiduk, D. M.; Holm, E. A.; Niezgoda, S. R. Perspectives on the Impact of Machine Learning, Deep Learning, and Artificial Intelligence on Materials, Processes, and Structures Engineering. *Integr Mater Manuf Innov* **2018**, *7* (3), 157–172. https://doi.org/10.1007/s40192-018-0117-8.

(13) Schmidt, J.; Marques, M. R. G.; Botti, S.; Marques, M. A. L. Recent Advances and Applications of Machine Learning in Solid-State Materials Science. *NPJ Comput Mater* **2019**, *5* (1). https://doi.org/10.1038/s41524-019-0221-0.

(14) Butler, K. T.; Davies, D. W.; Cartwright, H.; Isayev, O.; Walsh, A. Machine Learning for Molecular and Materials Science. *Nature* **2018**, *559* (7715), 547–555. https://doi.org/10.1038/s41586-018-0337-2.

(15) Kemp, R.; Cottrell, G. A.; Bhadeshia, H. K. D. H.; Odette, G. R.; Yamamoto, T.; Kishimoto, H. Neural-Network Analysis of Irradiation Hardening in Low-Activation Steels. *Journal of Nuclear Materials* **2006**, *348* (3), 311–328. https://doi.org/10.1016/j.jnucmat.2005.09.022.

(16) Cottrell, G. A.; Kemp, R.; Bhadeshia, H. K. D. H.; Odette, G. R.; Yamamoto, T. Neural Network Analysis of Charpy Transition Temperature of Irradiated Low-Activation Martensitic Steels. *Journal of Nuclear Materials* **2007**, *367-370 A* (SPEC. ISS.), 603–609. https://doi.org/10.1016/j.jnucmat.2007.03.103.

(17) Windsor, C.; Cottrell, G.; Kemp, R. Prediction of Yield Stress and Charpy Transition Temperature in Highly Neutron Irradiated Ferritic Steels. *Model Simul Mat Sci Eng* **2010**, *18* (5). https://doi.org/10.1088/0965-0393/18/5/055012.

(18) Long, S.; Zhao, M. Theoretical Study of GDM-SA-SVR Algorithm on RAFM Steel. *Artif Intell Rev* **2020**, *53* (6), 4601–4623. https://doi.org/10.1007/s10462-020-09803-y.

(19) Obraztsov, S. M.; Birzhevoi, G. A.; Konobeev, Y. V.; Pechenkin, V. A.; Rachkov, V. I. Neuronet Analysis of the Effect of Alloying Elements on the Radiation Embrittlement of VVÉR-440 Vessel Materials. *Atomic Energy* **2006**, *101* (5), 809–815. https://doi.org/10.1007/s10512-006-0173-6.

(20) Castin, N.; Malerba, L.; Chaouadi, R. Prediction of Radiation Induced Hardening of Reactor Pressure Vessel Steels Using Artificial Neural Networks. *Journal of Nuclear Materials* **2011**, *408* (1), 30–39. https://doi.org/10.1016/j.jnucmat.2010.10.039.

(21) Takamizawa, H.; Itoh, H.; Nishiyama, Y. Statistical Analysis Using the Bayesian Nonparametric Method for Irradiation Embrittlement of Reactor Pressure Vessels.

*Journal of Nuclear Materials* **2016**, *479*, 533–541. https://doi.org/10.1016/j.jnucmat.2016.07.035.

(22) Mathew, J.; Parfitt, D.; Wilford, K.; Riddle, N.; Alamaniotis, M.; Chroneos, A.; Fitzpatrick, M. E. Reactor Pressure Vessel Embrittlement: Insights from Neural Network Modelling. *Journal of Nuclear Materials* **2018**, *502*, 311–322. https://doi.org/10.1016/j.jnucmat.2018.02.027.

(23) Lee, G. G.; Kim, M. C.; Lee, B. S. Machine Learning Modeling of Irradiation Embrittlement in Low Alloy Steel of Nuclear Power Plants. *Nuclear Engineering and Technology* **2021**, *53* (12), 4022–4032. https://doi.org/10.1016/j.net.2021.06.014.

(24) Liu, Y. chen; Wu, H.; Mayeshiba, T.; Afflerbach, B.; Jacobs, R.; Perry, J.; George, J.; Cordell, J.; Xia, J.; Yuan, H.; Lorenson, A.; Wu, H.; Parker, M.; Doshi, F.; Politowicz, A.; Xiao, L.; Morgan, D.; Wells, P.; Almirall, N.; Yamamoto, T.; Odette, G. R. Machine Learning Predictions of Irradiation Embrittlement in Reactor Pressure Vessel Steels. *NPJ Comput Mater* **2022**, *8* (1), 1–11. https://doi.org/10.1038/s41524-022-00760-4.

(25) Ferreño, D.; Serrano, M.; Kirk, M.; Sainz-aja, J. A. Prediction of the Transition-Temperature Shift Using Machine Learning Algorithms and the Plotter Database. *Metals (Basel)* **2022**, *12* (2), 1–24. https://doi.org/10.3390/met12020186.

(26) Xu, C.; Liu, X.; Wang, H.; Li, Y.; Jia, W.; Qian, W.; Quan, Q.; Zhang, H.; Xue, F. A Study of Predicting Irradiation-Induced Transition Temperature Shift for RPV Steels with XGBoost Modeling. *Nuclear Engineering and Technology* **2021**, *53* (8), 2610–2615. https://doi.org/10.1016/j.net.2021.02.015.

(27) He, W. ke; Gong, S. yi; Yang, X.; Ma, Y.; Tong, Z. feng; Chen, T. Study on Irradiation Embrittlement Behavior of Reactor Pressure Vessels by Machine Learning Methods. *Ann Nucl Energy* **2023**, *192*. https://doi.org/10.1016/j.anucene.2023.109965.

(28) Liu, Y. -c.; Morgan, D.; Yamamoto, T.; Odette, G. R. Characterizing the Flux Effect on the Irradiation Embrittlement of Reactor Pressure Vessel Steels Using Machine Learning. *Acta Mater* **2023**, *256* (119144). https://doi.org/https://doi.org/10.1016/j.actamat.2023.119144.

(29) Morgan, D.; Pilania, G.; Couet, A.; Uberuaga, B. P.; Sun, C.; Li, J. Machine Learning in Nuclear Materials Research. *Curr Opin Solid State Mater Sci* **2022**, *26* (2), 100975. https://doi.org/10.1016/j.cossms.2021.100975.

(30) Bing, B.; Han, X.; Jia, L.; He, X.; Zhang, C.; Yang, W. Influence Analysis of Alloy Elements on Irradiation Embrittlement of RPV Steel Based on Deep Neural Network. *International Journal of Advanced Nuclear Reactor Design and Technology* **2023**. https://doi.org/10.1016/j.jandt.2023.03.002.

(31) Palmer, G.; Du, S.; Politowicz, A.; Emory, J. P.; Yang, X.; Gautam, A.; Gupta, G.; Li, Z.; Jacobs, R.; Morgan, D. Calibration after Bootstrap for Accurate Uncertainty Quantification in Regression Models. *NPJ Comput Mater* **2022**, *8* (1), 1–9. https://doi.org/10.1038/s41524-022-00794-8.

(32) G. R. Odette; T. Yamamoto; D. Klingensmith; D. Gragg; K. Fields; P. Wells; N. Almirall. *UCSB MRPG-RPV: 23-1 The UCSB IVAR, ATR-2 and BR2 Irradiation Experiments and the RPV Steel Hardening Database Used in Recent Machine Learning Studies*. https://doi.org/10.6084/m9.figshare.23304227.

(33)  Erickson, M.; Kirk, M. Use of Unirradiated Yield Strength as a Variable in Embrittlement Trend Forecasting to Better Inform D T 41J Predictions. **2022**.

(34)  Nanstad, R. K.; Almirall, N.; Wells, P.; Server, W. L.; Sokolov, M. A.; Long, E. J.; Odette, G. R. On High Fluence Irradiation Hardening of Nine RPV Surveillance Steels in the UCSB ATR-2 Experiment: Implications to Extended Life Embrittlment Predictions. *Submitted for review* **2022**.

(35)  Odette, G. R.; Lombrozo, P. M.; Wullaert, R. A. Relationship Between Irradiation Hardening and Embrittlement of Pressure Vessel Steels. *ASTM Special Technical Publication 870* **1985**, 840–860.

(36)  Lee, G. G.; Lee, Y.; Kwon, J. Relationship between Radiation Inducted Yield Strength Increment and Charpy Transition Temperature Shift in Reactor Pressure Vessel Steels of Korean Nuclear Power Plants. *Nuclear Engineering and Technology* **2012**, *44* (5), 543–550. https://doi.org/10.5516/NET.07.2011.022.

(37)  Nanstad, R. K.; Odette, G. R.; Almirall, N.; Robertson, J. P.; Server, W. L.; Yamamoto, T.; Well, P. *Effects of ATR-2 Irradiation to High Fluence on Nine RPV Surveillance Materials*; 2017.

(38)  Jacobs, R.; Mayeshiba, T.; Afflerbach, B.; Miles, L.; Williams, M.; Turner, M.; Finkel, R.; Morgan, D. The Materials Simulation Toolkit for Machine Learning (MAST-ML): An Automated Open Source Toolkit to Accelerate Data-Driven Materials Research. *Comput Mater Sci* **2020**, *176*, 109544. https://doi.org/10.1016/j.commatsci.2020.109544.

(39)  Pedregosa, F.; Varoquaux, G.; Gramfort, A.; Michel, V.; Thirion, B.; Grisel, O.; Blondel, M.; Prettenhofer, P.; Weiss, R.; Dubourg, V.; Vanderplas, J.; Passos, A.; Cournapeau, D.; Brucher, M.; Perrot, M.; Duchesnay, É. Scikit-Learn: Machine Learning in Python. *Journal of Machine Learning Research* **2011**, *12*.

(40)  Chollet, F. *Keras*. https://github.com/keras-team/keras.

(41)  Abadi, M.; Barham, P.; Chen, J.; Chen, Z.; Davis, A.; Dean, J.; Devin, M.; Ghemawat, S.; Irving, G.; Isard, M.; Kudlur, M.; Levenberg, J.; Monga, R.; Moore, S.; Murray, D. G.; Steiner, B.; Tucker, P.; Vasudevan, V.; Warden, P.; Wicke, M.; Yu, Y.; Zheng, X.; Brain, G. TensorFlow: A System for Large-Scale Machine Learning. In *12th USENIX Symposium on Operating Systems Design and Implementation (OSDI '16)*; 2016; pp 265–284. https://doi.org/10.1038/nn.3331.

(42)  Ferreño, D.; Kirk, M.; Serrano, M.; Sainz-Aja, J. A. Assessment of the Generalization Ability of the ASTM E900-15 Embrittlement Trend Curve by Means of Monte Carlo Cross-Validation. *Metals (Basel)* **2022**, *12* (3), 481. https://doi.org/10.3390/met12030481.

(43)  Lu, H. J.; Zou, N.; Jacobs, R.; Afflerbach, B.; Lu, X. G.; Morgan, D. Error Assessment and Optimal Cross-Validation Approaches in Machine Learning Applied to Impurity Diffusion. *Comput Mater Sci* **2019**, *169*, 109075. https://doi.org/10.1016/j.commatsci.2019.06.010.

(44)  Pernot, P. Validation of Uncertainty Quantification Metrics: A Primer Based on the Consistency and Adaptivity Concepts. **2023**.

(45)  Lundberg, S. M.; Allen, P. G.; Lee, S.-I. A Unified Approach to Interpreting Model Predictions. In *31st Conference on Neural Information Processing Systems (NIPS)*; 2017.

(46)  Odette, G. R.; Lucas, G. E. The Effect of Heat Treatment on Irradiation Hardening of Pressure Vessel Steels. In *Proceedings of the third international symposium on environmental degradation of materials in nuclear power systems*; 1988.

(47) Odette, G. R.; Lucas, G. E.; Klingensmith, D.; Wirth, B. D.; Gragg, D. Effects of Composition and Heat Treatment on Hardening and Embrittlement of Reactor Pressure Vessel Steels. In *NUREG/CR-6778* ; 2003.

(48) Odette, G. R.; Lucas, G. E. Recent Progress in Understanding Reactor Pressure Vessel Steel Embrittlement. *Radiation Effects and Defects in Solids* **1998**, *144* (1–4), 189–231. https://doi.org/10.1080/10420159808229676.

(49) E.D. Eason; G.R. Odette; R.K. Nanstad; Yamamoto, T. A Physically Based Correlation of Irradiation-Induced Transition Temperature Shifts for RPV Steels. *ORNL/TM-2006/530* **2007**.

(50) Odette *, G. R.; Yamamoto, T.; Klingensmith, D. On the Effect of Dose Rate on Irradiation Hardening of RPV Steels. *Philosophical Magazine* **2005**, *85* (4–7), 779–797. https://doi.org/10.1080/14786430412331319910.

(51) Odette, G. R.; Nanstad, R. K. Predictive Reactor Pressure Vessel Steel Irradiation Embrittlement Models: Issues and Opportunities. *Jom* **2009**, *61* (7), 17–23. https://doi.org/10.1007/s11837-009-0097-4.

(52) R. K. Nanstad; G. R. Odette; R. E. Stoller; T. Yamamoto. *Review of Draft NUREG Report on Technical Basis for Revision of Regulatory Guide 1.99, ORNL/NRC/LTR-08/03*; 2008.

(53) G. R. Odette; E. V. Mader; G. E. Lucas; W. J. Phythian; C. A. English. The Effect of Flux on the Irradiation Hardening of Pressure Vessel Steels. In *Effects of Radiation on Materials: 16th International Symposium*; 1994; p 373.

(54) Chaouadi, R.; Gérard, R. Neutron Flux and Annealing Effects on Irradiation Hardening of RPV Materials. *Journal of Nuclear Materials* **2011**, *418* (1–3), 137–142. https://doi.org/10.1016/j.jnucmat.2011.06.012.

(55) Kuleshova, E. A.; Zhuchkov, G. M.; Fedotova, S. V.; Maltsev, D. A.; Frolov, A. S.; Fedotov, I. V. Precipitation Kinetics of Radiation-Induced Ni-Mn-Si Phases in VVER-1000 Reactor Pressure Vessel Steels under Low and High Flux Irradiation. *Journal of Nuclear Materials* **2021**, *553*, 153091. https://doi.org/10.1016/j.jnucmat.2021.153091.

(56) Dohi, K.; Onchi, T.; Kano, F.; Fukuya, K.; Narui, M.; Kayano, H. Effect of Neutron Flux on Low Temperature Irradiation Embrittlement of Reactor Pressure Vessel Steel. *Journal of Nuclear Materials* **1999**, *265* (1–2), 78–90. https://doi.org/10.1016/S0022-3115(98)00611-4.

(57) Wagner, A.; Bergner, F.; Chaouadi, R.; Hein, H.; Hernández-Mayoral, M.; Serrano, M.; Ulbricht, A.; Altstadt, E. Effect of Neutron Flux on the Characteristics of Irradiation-Induced Nanofeatures and Hardening in Pressure Vessel Steels. *Acta Mater* **2016**, *104*, 131–142. https://doi.org/10.1016/j.actamat.2015.11.027.

(58) Bergner, F.; Ulbricht, A.; Hein, H.; Kammel, M. Flux Dependence of Cluster Formation in Neutron-Irradiated Weld Material. *Journal of Physics: Condensed Matter* **2008**, *20* (10), 104262. https://doi.org/10.1088/0953-8984/20/10/104262.

(59) Kempf, R.; Troiani, H.; Fortis, A. M. Effect of Lead Factors on the Embrittlement of RPV SA-508 Cl 3 Steel. *Journal of Nuclear Materials* **2013**, *434* (1–3), 411–416. https://doi.org/10.1016/j.jnucmat.2012.12.004.

(60) Ulbricht, A.; Hernández-Mayoral, M.; Oñorbe, E.; Etienne, A.; Radiguet, B.; Hirschmann, E.; Wagner, A.; Hein, H.; Bergner, F. Effect of Neutron Flux on an Irradiation-Induced

Microstructure and Hardening of Reactor Pressure Vessel Steels. *Metals (Basel)* **2022**, *12* (3), 369. https://doi.org/10.3390/met12030369.